\documentclass[fleqn,usenatbib]{mnras}

\usepackage{newtxtext,newtxmath}

\usepackage[T1]{fontenc}
\usepackage{ae,aecompl}

\usepackage{graphicx}	% Including figure files
\usepackage{amsmath}	% Advanced maths commands
\usepackage[page]{appendix}

\usepackage{bm}
\usepackage{tabularx}
\usepackage{subcaption}
\usepackage{lastpage}

\usepackage{soul}
\renewcommand\hl[1]{#1} % <<<<<<<<<<<<<<<<<<<<<<<<<<<<<<<<<<<<<

\usepackage{xcolor}
\usepackage[normalem]{ulem} % use normalem to protect \emph

\usepackage[section]{placeins}

\usepackage{float}
\title[CR and ionizing radiation effects on CGM]{Effect of cosmic rays and ionizing radiation on observational ultraviolet plasma diagnostics in the circumgalactic medium}

\author[Holguin et al..]{
F. Holguin,$^{1,2}$\thanks{E-mail: fholgui1@jhu.edu(FH)}
R. Farber,$^{3,4}$
J. Werk$^{5}$
\\
$^{1}$Johns Hopkins University Applied Physics Laboratory, Laurel, MD 2128, USA\\
$^{2}$Whitling School of Engineering, Johns Hopkins University, Baltimore, MD 21218, USA\\
$^{3}$Max Planck Institut für Astrophysik, Karl-Schwarzschild-Straße 1, 85748 Garching bei München, Germany\\
$^{4}$Department of Physics, Purdue University Fort Wayne, Fort Wayne, IN 46805, USA\\
$^{5}$University of Washington, Seattle, WA 98195, USA; email: jwerk@uw.edu}

\date{Accepted XXX. Received YYY; in original form ZZZ}

\pubyear{2024}

\begin{document}
\label{firstpage}
\pagerange{\pageref{firstpage}--\pageref{lastpage}}
\maketitle

% Abstract of the paper
\begin{abstract}
The relevance of some galactic feedback mechanisms, in particular cosmic ray feedback and the hydrogen ionizing radiation field, has been challenging to definitively describe in a galactic context, especially far outside the galaxy in the circumgalactic medium (CGM). Theoretical and observational uncertainties prevent conclusive interpretations of multiphase CGM properties derived from ultraviolet (UV) diagnostics. We conduct three dimensional magnetohydrodynamic simulations of a section of a galactic disk with star formation and feedback, including radiative heating from stars, a UV background, and cosmic ray feedback. We utilize the temperature phases present in our simulations to generate Cloudy models to derive spatially and temporally varying synthetic UV diagnostics. We find that radiative effects \hl{without additional heating mechanisms} are not able to produce synthetic diagnostics in the observed ranges. For low cosmic ray diffusivity $\kappa_{\rm{cr}}=10^{28} \rm{cm}^2 \rm{s}^{-1}$, cosmic ray streaming heating in the outflow helps our synthetic line ratios roughly match observed ranges by producing transitional temperature gas ($T \sim 10^{5-6}$ K). High cosmic ray diffusivity $\kappa_{\rm{cr}}=10^{29} \rm{cm}^2 \rm{s}^{-1}$, with or without cosmic ray streaming heating, produced transitional temperature gas. The key parameter controlling the production of this gas phase remains unclear, as the different star formation history and outflow evolution itself influences these diagnostics. Our work demonstrates the use of UV plasma diagnostics to differentiate between galactic/circumgalactic feedback models.
\end{abstract}

% Select between one and six entries from the list of approved keywords.
% Don't make up new ones.

\section{Introduction}

Simulations of the galactic and cosmological environment have greatly improved over the last decade. More detailed models of stellar and radiative feedback, along with increased computational power, have allowed modern three dimensional galactic \citep[e.g.,][]{guedes2011forming,hopkins2012,rosdahl2013ramses} and full cosmological-scale \citep[e.g.,][]{vogelsberger2014introducing,schaye2015eagle,pillepich2018simulating,dave2019simba} models to reproduce many global properties observed in real galaxies, such as the efficiency of stellar feedback for sub-Milky Way and Milky Way mass halos \citep[][]{hopkins2012stellar}, SFR variability \citep[][]{orr2017stacked, sparre2017star}, or gas velocity dispersion \cite[e.g.][]{orr2020swirls}. Previously neglected processes have also been implemented, such as turbulence \citep[][]{schmidt2014cosmological}, magnetohydrodynamics (MHD) \cite[e.g.][]{wang2009magnetohydrodynamic,dolag2009mhd, hopkins2018fire,steinwandel2022driving}, cosmic rays (CRs) \citep[e.g.][]{booth2013,salem2014cosmic,hopkins2020but,rathjen2021silcc,girichidis2023spectrally}, and radiative feedback \citep[][]{rosdahl2013ramses,wibking2018radiation,kannan2022introducing,katz2022ramses,bhagwat2023spice}. These galactic simulations facilitate better interpretations of astronomical measurements from real galaxies as they are a `universe-in-a-box' allowing us to isolate and closely examine the effects of individual physical processes.

Here we focus on Milky Way-like galaxies, where stellar feedback (including cosmic rays) is likely to play a role in galactic evolution \cite[i.e.,][]{hopkins2012, hopkins2014galaxies, ruszkowski2017global, habegger2022impact, butsky2023constraining}. Radiation and CR phenomena are similar in that small-scale physics, much below the resolution scale of a galactic simulation, plays a key role in global transport. For example, two opposing limits--the free-streaming and diffusive limits--must be adequately modeled \cite[e.g.,][]{jiang2018new, thomas2019cosmic, hopkins2022consistent}. The stellar radiation field likely extends well past the galactic disk and into the inner CGM up to roughly 0.1-0.2 $R_{\rm{vir}}$ for Milky Way-like, low redshift galaxies \citep[][]{holguin2023contribution}. Therefore, it is important to consider the influence of this spatially and temporally varying field on the ionization state in the inner CGM. Three dimensional radiative transfer is incredibly computationally expensive as the radiation field acts non-locally.

CR physics is challenging to model as there are large uncertainties in the underlying assumptions about the interaction of CR and the magnetized environment. CRs are known to have a galactic confinement time much larger than the free streaming time and their arrival distribution measured at the Earth for Tev-PeV cosmic rays is measured to be isotropic to 10$^{-4}$ \cite[][]{nagashima1998galactic}. CR confinement is primarily described by two ideas, extrinsic turbulence and self-confinement \citep[see][]{zweibel2013microphysics}. Within both of these descriptions, many models have been developed and even implemented in galactic simulations \citep[see][]{zweibel2013microphysics, zweibel2017basis, thomas2019cosmic}. However, distinguishing between these models based on observational evidence of CRs, such as the total column density of matter traversed by CRs (`grammage') \citep[][]{strong2007cosmic}, radio synchrotron (e.g., the CHANG-ES radio halo survey \citealt[][]{irwin2012continuum}) or gamma-rays \citep[e.g.,][]{ackermann2012fermi} has been difficult.  \citet[][]{hopkins2021testing} compares a wide range of CR transport parameters in dwarf and MW-type galaxy simulations and identifies a set of models that are consistent with observational evidence. These allowed models of CR transport, including more physically motivated ones, still have significant uncertainties. They require some fine-tuning, tend to produce similar effects on galaxy and inner-CGM (<10 kpc) properties, and are not well-constrained observationally farther out into the CGM \citep[][]{hopkins2021effects}. These models are also problematic with regards to constraints from the CR spectrum \citep[][]{hopkins2021standard}. The large uncertainty of CR transport motivates the introduction of additional constraints. 

Ultraviolet (UV) ion diagnostics contain a wealth of information about the plasma state of the CGM from the cold, neutral medium to the hot, ionized galactic corona \citep[see][and references therein]{tumlinson2017}. These diagnostics have highlighted the CGM as a significant reservoir of baryons existing in a multiphase medium. The exact properties of the plasma that produces some of these observed diagnostics is uncertain. Untangling the contributions from unresolved gas components requires careful statistical analysis of the spectra, including accounting for multiple, co-spatial components \citep[][]{cooper2021cosmic}. In particular, there is evidence based on the ionization state of metals that there exists a $T \sim 10^{5} $ K phase, in between classical ISM and coronal stable phases \citep[][]{mckee1977theory, wolfire2003neutral}. The existence of a $10^5$\,K phase is perplexing as it sits near the peak of the (solar metallicity) cooling curve \citep[][]{sutherlanddopita1993cooling}. Unfortunately, modeling of this (thermally) unstable phase is complicated by degeneracies in modeling for local ionization equilibirum; similar diagnostic signals can occur from collisionally ionized $T \sim 10^{5}$ K and photoionized $T \sim 10^{4}$ K plasmas. The collisional ionization explanations typically invoke additional heating physics, such as (radiative) turbulent mixing layers \cite[e.g.][]{begelman1990,kwak2010numerical, ji2019simulations} or collisionless heating from cosmic ray streaming \cite[][]{wiener2013cosmicheating}. On the other hand, ionization states observed may be from photoionization \cite[][]{muzahid2015extreme}. In addition to the heating, CRs can also influence the properties of the multiphase plasma \cite[e.g.][]{ji2020properties,butsky2020impact} via their pressure, potentially affecting these diagnostics even further. 

Galactic computational simulations can aid in untangling these degeneracies. CR feedback models have become progressively more detailed \citep[i.e.][]{werhahn2021cosmic2, werhahn2021cosmic3} and interest in tackling these interdependent multiphase and multiphysics questions is growing \cite[][]{faucher2023key}. Several simulations now include three-dimensional radiative transfer at run time while remaining computational feasible \citep[][]{rathjen2021silcc, kim2022photochemistry,katz2022ramses}. At the same time, UV diagnostic data will continue to improve, such as with the James Web space telescope and 30-meter class ground-based telescopes. This data will illuminate CGM spectroscopic properties up to cosmic noon $z \sim 2$. In order to take full advantage of the increased quantity and quality of observed data, it is crucial for simulations to make predictions aligned with these diagnostics \citep[e.g.][]{rathjen2023optical}. 

Our goal is to provide a proof-of-concept of CGM UV absorption-line diagnostic trends derived from three-dimensional MHD simulations, focusing on radiative heating and CR feedback. In Section \ref{sec:methods}, we describe the MHD simulation domain and feedback implementations. In Section \ref{simulation_results} we describe results from updated parallel-plane elongated slab simulations from \citet[][]{holguin2019role} with simplified radiative heating, improved supernova (SN) feedback, and CR collisional losses. In these six simulations, we explore several limiting cases in the parameter space of CR and radiation heating feedback. In Section \ref{postprocessing_section} we describe a parallel-plane post-processing approximation, including the stellar and metagalactic radiation field, that can estimate the ionization state of the plasma throughout the domain. We also examine theoretical expectations for CR and radiation feedback effects on UV diagnostics tracing gas phases between $T=10^{4}$ K and $T=10^{6}$ K. In Sections \ref{results} and \ref{discussion}, we describe resulting synthetic diagnostic predictions and broadly compare them to typical observed values. Finally, in Section \ref{conclusions} we provide concluding remarks.

\section{MHD simulation methods}
\label{sec:methods}
The simulations we run are similar to those from \citet{holguin2019role}, with updated stellar feedback methods and additional radiative heating and CR loss models. The simulations are run with the adaptive mesh refinement MHD code FLASH 4.2 \citep{fryxell2000flash, dubey2008introduction} using a directionally unsplit staggered mesh solver \citep{lee2009unsplit, lee2013solution}, including CR transport physics \citep{yang2012fermi, ruszkowski2017global, farber2018impact} in an elongated box geometry with dimensions $2 \times 2 \times 40 \ \rm{kpc}^3$ \cite[][]{hill2012vertical, walch2015silcc}.

We solve a coupled MHD-CR two-fluid model \citep{salem2014cosmic, ruszkowski2017global} composed of thermal gas and ultrarelativistic CR fluid with adiabatic indices $\gamma = 5/3$ and $\gamma_{\rm{cr}} = 4/3$ respectively. The CR fluid represents a single energy channel of typical galactic CRs with mean CR Lorentz factor $\gamma_{\rm{rel}} = 3$ and momentum distribution slope $n=4.5$. Both hadronic and Coulomb CR losses are included \citep[][]{guo2008feedback}. CR transport includes advection, dynamical coupling with gas, and \hl{diffusing relative to the magnetic field lines with spatially and temporallay constant diffusivities}. Heating of the gas occurs due to radiative heating from both non-ionizing and ionizing UV stellar spectrum, CR streaming and hadronic losses, and supernova feedback.  Cooling of gas occurs via radiative cooling. Gas dynamics is impacted by self-gravity, stellar momentum feedback, and dynamical coupling with CRs. Star formation occurs according to \citep[][]{cen1992galaxy}, resulting in the creation of a stellar population particle and equal mass loss of gas (see \ref{sfeed} for more details). The following equations summarize the model:

\begin{align}
\frac{\partial \rho}{\partial t} + \bm{\nabla} \cdot ( \rho \bm{u}_g )   =  -\dot{m}_{\textrm{form}} + f_{\ast} \dot{m}_{\textrm{feed}}
\label{eom1}
\end{align}

\begin{equation}
\frac{\partial \rho \bm u_g}{\partial t} + \bm{\nabla} \cdot \left(  \rho \bm{u}_g \bm{u}_g - \frac{\bm{B}\bm{B}}{4\pi} \right) + \bm{\nabla} p_{\textrm{tot}} = \rho \bm{g} + \dot{p}_{\rm SN} 
\label{eom2}
\end{equation}

\begin{equation}
\frac{\partial \bm{B}}{\partial t} - \bm{\nabla} \times (\bm{u}_g \times \bm{B} ) = 0
\label{eom3}
\end{equation}

\begin{equation}
\begin{split}
\frac{\partial e}{\partial t} + & \bm{\nabla} \cdot \left[ (e + p_{\textrm{tot}}) \bm{u}_g - \frac{\bm{B}(\bm{B} \cdot \bm{u}_g  )}{4\pi}    \right]  = \rho \bm{u}_g \cdot \bm{g}  \\
&- \nabla \cdot \bm{F}_{\rm cr}  - C + H_{\rm SN} + H_{\rm{rad}} + \Gamma_{\rm{cr, gas}}
\end{split}
\label{eom4}
\end{equation}

\begin{equation}
\begin{split}
\frac{\partial e_{\rm cr} }{\partial t} + \bm{\nabla} \cdot(e_{\rm cr} \bm{u}_g ) = & -p_{\rm cr}  \nabla \cdot \bm{u}_g - H_{\rm cr} + \Gamma_{\rm{cr}} + H_{\rm SN} \\
& - \nabla \cdot (\bm{\kappa}_{\rm{cr}} \cdot \nabla e_{\rm{cr}})
\end{split}
\label{eom5}
\end{equation}

\begin{equation}
\Delta \phi = 4\pi G \rho_b
\label{eom6}
\end{equation}

\noindent where $\rho$ is the gas density, $\rho_b$ is the total baryon density including both the gas and stars, $\dot{m}_{\textrm{form}}$ is the density sink from stellar population particle formation, $f_{\ast} \dot{m}_{\textrm{feed}}$ represents the gas density source from stellar feedback (see Section \ref{sfeed}), $\bm{u}_g$ is the gas velocity, $\bm{B}$ is the magnetic field, $G$ is the gravitational constant, $\phi$ is the gas gravitational potential, $\bm{g} = - \bm{\nabla} \phi + \bm{g}_{\textrm{NFW}}$ is the gravitational acceleration (the sum of gas self-gravity, stellar particle, and halo dark matter contributions to the gravitational acceleration, described in Section \ref{grav}) where $\bm{g}_{\textrm{NFW}}$ is the gravity from the Navarro-Frenk-White (NFW) potential, $p_{\textrm{tot}}$ is the sum of gas ($p_{\rm th}$), magnetic, and CR ($p_{\rm cr}$) pressures, $\dot{p}_{\rm SN}$ is the momentum injection due to stellar winds and SNe.  Furthermore, $e = \rho \bm{u}^2_g + e_g + e_{\rm cr} + B^2 / 8\pi$ is the total energy density per volume (the sum of gas, CR, and magnetic components, respectively), $C$ is the radiative cooling rate per unit volume,  $H_{\rm{rad}}$ is the radiative heating rate per unit volume, $\Gamma_{\rm{cr}}$ is the CR energy loss rate per unit volume, $\Gamma_{\rm{cr, gas}}$ is the energy rate gained by the gas from the CR losses, $\kappa_{\rm{cr}}$ is the CR diffusion coefficient, and $H_{\rm SN}$ is the supernova heating rate per volume.

\subsection{Gravity} \label{grav}
The gravitational acceleration has three components: baryons, stellar particles, and the dark matter halo. Self-gravity from baryons is computed by solving the Poisson equation with the Barnes-Hut tree solver \citep{barnes1986hierarchical}, which is implemend in FLASH by \citet{wunsch2018tree}. For the dark matter halo gravity modeled as an NFW potential \citep[][]{navarro1997universal}, we include the vertical component of gravity as the domain is thin and much smaller than the halo. The halo gravitational acceleration is 

\begin{equation}
g_{\textrm{NFW}} (z) = - \frac{G M_{200}}{|z|^3} \frac{\textrm{ln}(1+ x)    - x/(1+x)  }{\textrm{ln}(1+c) - c/(1+c)}
\end{equation}

\noindent where $G$ is the gravitational constant, $M_{200}$ is the halo virial mass, $z$ is the height above the mid-plane, $x = |z| c / r_{200}$, $c$ is the halo concentration parameter, and $r_{200}$ is the virial radius. Table \ref{sim_param_table} summarizes our choices of these parameters.

\subsection{Star formation and supernova feedback} \label{sfeed}

Following the approach of \citet{cen1992galaxy} (see also \citealt{tasker2006simulating,bryan2014enzo,salem2014cosmic, li2015cooling}), the creation of a stellar population particles occurs when all of the following conditions are simultaneously met: (i) gas number density exceeds $10 \ \textrm{cm}^{-3}$ \citep{gnedin2011environmental, agertz2013toward}; (ii) the cell mass exceeds the local Jeans mass; (iii) the divergence of the gas velocity $\bm{u}_g$ is negative, $\nabla \cdot \bm{u}_g < 0$; (iv) gas temperature reaches the floor of the cooling function or the cooling time becomes shorter than the dynamical time $t_{\textrm{dyn}} = \sqrt{3\pi / (32G\rho_b)}$. The particle is formed with a velocity equal to the surrounding gas, mass $m_{\ast} = \epsilon_{\textrm{SF}} (dt/t_{\textrm{dyn}}) \rho dx^3 $, where $\epsilon_{\textrm{SF}} = 0.05$ is the star formation efficiency, $dt$ is the timestep, and  $dx$ is the local cell size.  A mass equal to the particle mass is removed from the gas. We set a minimum particle mass $m_{*,\textrm{min}} = 10^5 M_{\odot}$ in order to manage the number of particles formed, but we still permit particles with $m_{\ast} < m_{*,\textrm{min}}$ to form with a probability $m_{\ast}/m_{*,\textrm{min}}$ and mass $m_{\ast} = 0.8 \rho dx^3$.

In order to model feedback due to stellar winds and SNe, we inject gas thermal energy and momentum, as well as CRs energy. The gas and CR energy is injected into the particle's local cell, while the momentum is injected into the immediately surrounding cells as in \citet{farber2022stress}. \footnote{\citet{holguin2019role} did not include this addition of momentum feedback.} 
\noindent
The inclusion of momentum feedback is essential when the resolution does not sufficiently resolve the cooling radius, as the energy injected can be quickly radiated away and fixes such as delayed cooling over predict the energy and momentum at late-times in SN evolution \citep{martizzi2015supernova}. The full details of the momentum feedback are found in \citet{farber2022stress} and we briefly summarize here. We use the following model for SNe evolution in an inhomogeneous medium, found in Eq. 12 of \citet{martizzi2015supernova}, where we assume solar metalicity:

\begin{equation}
\begin{split}
\alpha  =& \  -11 \left( \frac{n_{\rm{H}}}{100 \rm{cm}^{-3}} \right)^{0.114} \\
R_{\rm{c}}  =& \  6.3 \ \rm{pc} \left( \frac{n_{\rm{H}}}{100 \rm{cm}^{-3}} \right)^{-0.42} \\
R_{\rm{r}}  =& \  9.2 \ \rm{pc} \left( \frac{n_{\rm{H}}}{100 \rm{cm}^{-3}} \right)^{-0.44} \\
R_{\rm{0}}  =& \  2.4 \ \rm{pc} \left( \frac{n_{\rm{H}}}{100 \rm{cm}^{-3}} \right)^{-0.35} \\
R_{\rm{b}}  =& \  8.0 \ \rm{pc} \left( \frac{n_{\rm{H}}}{100 \rm{cm}^{-3}} \right)^{-0.46}.
\end{split}
\label{momentum1}
\end{equation}

The momentum $P_{\rm{r}}(R)$ deposition at a particular distance $R$ from center of the cell containing a stellar particle is written as 

\begin{equation}
\begin{split}
P_{\rm{r}}(R)/P_{0} = \left( \frac{R}{R_{0}} \right)^{1.5} \Theta(R_{\rm{b}} - R) + \left( \frac{R_{\rm{b}}}{R} \right) \Theta(R - R_{\rm{b}} )
\end{split}
\end{equation}
where $\Theta(x)$ is the Heaviside step function, $P_0 = f_{*,\rm{mom}} \sqrt{2 \phi \epsilon_{\rm{SN}} N_{\rm{SN}} 0.9 M_{\rm{ej,0}} }$. The parameter $f_{*,\rm{mom}} = 5$ represents a boost to the momentum received by the surrounding gas \citep[][]{agertz2013toward, semenov2018galaxies}, to compensate for advection errors, clustering of supernovae \citep[][]{gentry2017enhanced, gentry2020momentum} and backreaction of cosmic rays \citep[][]{diesing2018effect}.

\subsection{Radiative cooling} \label{cool}
Radiative cooling is calcuated using the Townsend cooling method \citep{townsend2009exact, zhu2017gas}, updated from \citet{farber2018impact, farber2022stress} to include net heating. We approximate  the radiative cooling functions from \citet{dalgarno1972heating} and \citet{raymond1976radiative} with a piecewise power law cooling function $\Lambda(T)$ in units of $\textrm{cm}^3 \textrm{s}^{-1}$ as

\begin{equation}
\Lambda(T) = \left\{\begin{aligned}  
    &0                                 &&\mbox{if                            $T$ $<$             300} \\
    &2.2380\times 10^{-32}T^{2.0  } &&\mbox{if               300 $\leq$ $T$ $<$            2000} \\
    &1.0012 \times 10^{-30}T^{1.5} &&\mbox{if              2000 $\leq$ $T$ $<$            8000} \\
    &4.6240 \times 10^{-36}T^{2.867} &&\mbox{if              8000 $\leq$ $T$ $<          10^{5}$} \\
    &1.7800 \times 10^{-18}T^{-0.65} &&\mbox{if          $10^{5}$ $\leq$ $T$ $< 4 \times 10^{7}$} \\
    &3.2217 \times 10^{-27}T^{0.5} &&\mbox{if $4 \times 10^{7}$ $\leq$ $T$, }\end{aligned}\right. 
\label{cooling}
\end{equation}

\noindent where $T$ is the gas temperature in K. This cooling function is appropriate for a gas of solar abundance, which is completely ionized at $T= 8000$ K.

\subsection{Radiative heating} \label{radheat}
Radiative heating from stellar sources is calculated using a temporally-varying parallel-plane model, similar to \citet{kim2017three}, including from both FUV (6 eV < $h\nu$ < 13.6 eV, far ultraviolet) and ionizing (or EUV, $h\nu > 13.6$ eV, extreme ultraviolet) spectrum ranges. Photoelectric heating occurs when FUV radiation interacts and frees electrons from interstellar dust grains \citep[]{weingartnerdraine2001photoelectric}. Photoionization heating occurs when ionizing radiation frees electrons from an atom, particularly hydrogen, and the electron eventually recombines \citep[][]{osterbrock2006astrophysics}.

\begin{figure}
    \centering
    \includegraphics[width=0.45\textwidth]{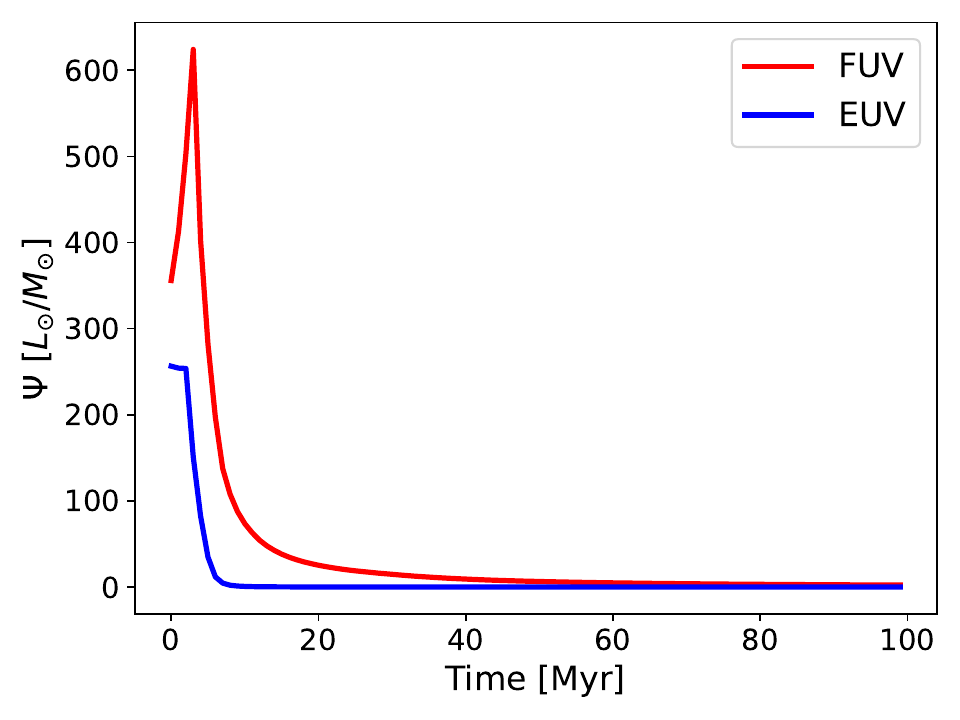}
        \caption{Luminosity-to-mass ratio ($L_{\odot}$  per $M_{\odot}$) from Starburst99 in the FUV and EUV wavelength ranges as a function of stellar population age. Using these data points, we assign a a luminosity to each stellar population particle, and the total luminosity is used to calculate the radiative heating rates in Eqs. \ref{peheat} and \ref{piheat}.  }
    \label{psiSB99}
    
\end{figure}

The radiative heating rates in those two ranges $H_{\rm{rad, FUV}}$ and $H_{\rm{rad, EUV}}$, are proportional to the total stellar population particle intensities $L_{\rm{FUV}}$ and $L_{\rm{EUV}}$ respectively, at a given simulation time. Due to the parallel-plane approximation, the position of the particles does not matter. We calculate the individual stellar particle intensities using STARBURST99 \citep{leitherer1999star}, which gives a spectrum (erg $\rm{s}^{-1} \rm{Hz}^{-1} M_{\odot}^{-1}$) of a stellar population at a fixed mass $M$ and age $t_{\rm{p}}$. We integrate the spectrum in the FUV and EUV ranges to derive the time-dependent luminosity-to-mass ratio $\Psi(t_{\rm{p}})$ shown in Figure \ref{psiSB99}. Since the hydrogen photoionization cross section drops rapidly above $13.6$ eV, we consider the EUV range to be 13.6 eV < $h\nu$ < 27.2 eV.

We calculate the total luminosity in each band by summing the luminosity of each particle, given its age and mass from $\Psi$. Radiative heating occurs only in neutral gas. \cite{kim2017three} models the transition between neutral and ionized gas regimes with a temperature dependent mean molecular weight interpolation, following results from \citet{sutherlanddopita1993cooling}. We instead use sharp transition between neutral and ionized regimes at a temperature $T_{\rm{ion}} = 1.5 \times 10^{4}$ K. This transition temperature is also used to determine the regime of CR losses. We use the photoelectric heating rate $\Gamma_{\rm{pe}}$ from \citet{kim2017three}, assuming a 4 $\rm{kpc}^2$ cross section for our parallel-plane domain. 

\begin{equation}
\begin{split}
H_{\rm{rad, FUV}} \sim & \ \Gamma_{\rm{pe,0}} \ n_{\rm{H}}  \left( \frac{\Sigma_{\rm{FUV}}}{\Sigma_{\rm{FUV,0}}} + 0.0024 \right)   \\
\sim & \ 2 \times 10^{-26} \left( \frac{n_{\rm{H}}}{\rm{cm}^{-3}} \right) \left( \frac{1}{2.8}  \left(   \frac{L_{\rm{FUV}}}{10^7 L_{\odot}}    \right)   + 0.0024     \right)    \ \rm{erg}\ \rm{s}^{-1}\rm{cm}^{-3}
\end{split}
\label{peheat}
\end{equation}
where $\Sigma_{\rm{\rm{FUV}}} = L_{\rm{FUV}}/ (l_{\rm x} l_{\rm y})$ and $l_{\rm x} = l_{\rm y} = 2 \ \rm{kpc}$.

The photoionization heating rate for hydrogen \citep{osterbrock2006astrophysics} is given by 
\begin{equation}
H_{\rm{rad, EUV}} = n_{\rm{H}} \int_{\nu_0}^{\infty} \frac{4\pi J_{\nu}}{h\nu} \sigma_{\nu} h(\nu - \nu_0) \rm{d}\nu,
\label{piheat}
\end{equation}
where $n_{\rm{H}}$ is the neutral hydrogen number density, $h\nu_{0} \approx 13.6 \ \rm{eV} \  = 1 \ \rm{Ryd}$  is the hydrogen ionization energy , $J_{\nu}$ is the specific intensity, $\sigma_{\nu}$ is the frequency-dependent ionization cross section of neutral hydrogen. We integrate in the range $h\nu$ = [1 Ryd, 2 Ryd] and assume $J_{\nu}$ is roughly constant over this interval. For a slab geometry with attenuating gas, the mean intensity is given by $4\pi J_{\rm{EUV}}$ = $\Sigma_{\rm{EUV}}(1 - E_{2}(\tau_{\perp}/2))/\tau_{\perp} = \Sigma_{EUV} A(z)$ \citep{ostriker2010}, where A(z) is the attenuation factor accounting for absorption of ionizing radiation by neutral hydrogen as a function of height $z$ above the galactic midplane, $E_2$ is the second exponential integral, $\tau_{\perp} = \kappa_{\rm{ion}} \Sigma(z)_{n}$, $\Sigma_{\rm{n,gas}}$ is the neutral column density from the midplane to a given $z$. All together $J_{\nu} \sim \frac{J_{\rm{EUV}}}{\Delta \nu} \sim \frac{L_{\rm{EUV}}}{\Delta \nu  \ 4 \ \rm{kpc}^2} $ for our geometry. The heating rate is calculated as follows

\begin{equation}
\begin{split}
H_{\rm{ion}} = & \ n_{\rm{H}} \int \frac{4\pi J_{\nu}}{h\nu} \sigma_{\nu} h(\nu - \nu_0) \rm{d}\nu  \\
\sim & \ n_{\rm{H}}     \left( \frac{J_{\rm{EUV}}}{\Delta \nu} \right) \ A(z) \ \int 4\pi \sigma_{\nu} \frac{h(\nu - \nu_0)}{h\nu} \rm{d}\nu \\
\sim & \ 5\times 10^{-22} \left( \frac{n_{\rm{H}}}{\rm{cm}^{-3}} \right) \left(\frac{L_{\rm{EUV}}      }{10^{7} L \odot} \right) \ A(z) \ \ \rm{erg}\ \rm{s}^{-1}\rm{cm}^{-3} \ \\
 \sim & 10^{-26} \left( \frac{n_{\rm{H}}}{\rm{cm}^{-3}} \right) \left(\frac{L_{\rm{EUV}}      }{10^{7} L_{\odot}} \right) \left(  \frac{ \Sigma_{\rm{n}} (z) }{50 M\odot/\rm{pc}^2} \right)^{-1}   \ \rm{erg}\ \rm{s}^{-1}\rm{cm}^{-3}
\end{split}
\label{piheatatten}
\end{equation}

The total radiative heating is $H_{\rm{rad}} (n_{\rm{H}}, L_{\rm{FUV}}, L_{\rm{EUV}}, z) = H_{\rm{rad, FUV}} + H_{\rm{rad, EUV}}$. The overall net energy recieved by gas is $\left( H_{\rm{rad}}  - \Lambda \right)$, which can take a positive or negative value. %The addition of net heating requires an modification of the Townsend cooling code. %Appendix \ref{ap_cool} describes the updated algorithm and tests.

We do not include dynamical coupling between the radiation field and gas. The FUV portion of the stellar spectrum exerts pressure on the dust, which is coupled to the gas. Generally, for winds driven by radiation pressure to be efficient, the stellar flux must be near the dust Eddington limit, which is usually the case in in high redshift and/or star bursting galactic environments \citep[][]{zhang2018review}. In both observed and simulated Milky Way-like galaxies, the focus of our work, other parts of stellar feedback likely dominate the acceleration of galactic winds \citep[][]{andrews2011assessing,hopkins2012stellar}.

\subsection{Cosmic ray physics} \label{crphys}

CR coupling to gas via advection and diffusive transport follows a similar model as in \citet{farber2018impact}. We use a constant value $\kappa_{\rm{cr}}$ for the CR diffusivity parallel to magnetic field lines. \footnote{We do not use the CR streaming model from \citet[][]{ruszkowski2017global, holguin2019role}.} As CRs travel through the ISM, they interact inelastically with particles in the ISM. We include both hadronic and Coulomb losses with a model from \citet{guo2008feedback}, as implemented in \citet{chan2019cosmic}, modeling $\Gamma_{\rm{cr}}$, the energy loss rate of the CRs, and $\Gamma_{\rm{cr,gas}}$, the amount of the CR loss rate that is absorbed by the gas (with the rest of the energy lost as gamma rays):

\begin{equation}
\begin{split}
\Gamma_{\rm{cr}} = \ - 5.86 \times 10^{-16} \ (1 + 0.282 \ x_e) \left( \frac{n_{\rm{n}}}{\rm{cm}^{-3}} \right) \left( \frac{e_{\rm{cr}}}{\rm{erg} \ \rm{cm}^{-3}} \right) \rm{erg} \ \rm{cm}^{-3} \rm{s}^{-1} \\
\Gamma_{\rm{cr,gas}} = \ 0.98 \times 10^{-16} \ (1 + 1.7 \ x_e) \left( \frac{n_{\rm{n}}}{\rm{cm}^{-3}} \right) \left( \frac{e_{\rm{cr}}}{\rm{erg} \ \rm{cm}^{-3}} \right) \rm{erg} \ \rm{cm}^{-3} \rm{s}^{-1}
\end{split}
\label{crloss_collision}
\end{equation}
where $n_{\rm{n}}$ is the number of nucleons, $x_e$ is the number of free electrons per nucleon, and $e_{\rm{cr}}$ is the CR energy density. %Appendix \ref{ap_cr} further describes the algorithm implementation and tests of its validity.

% mention Hcr from confinement model, and why we use it here
In the self-confinement model of CR transport \citep[][]{zweibel2013microphysics, zweibel2017basis}, CR streaming along the magnetic field lines is subject to the streaming instability, resulting in the generation of Alfv\`en waves. The subsequent wave-particle interactions between the population of CRs and  Alfv\`en waves limits the bulk CR transport speed to the Alfv\`en speed. This process also leads to energy transfer from the CRs to the thermal gas via collisionless heating, according to 

\begin{equation}
H_{\rm{cr}} = - v_{\rm{A}} \cdot \nabla P_{\rm{cr}},
\end{equation}
where $v_{\rm{A}}$ is the Alfv\`en speed and $P_{\rm{cr}}$ is the CR pressure \citep[][]{wiener2013cosmicheating}. We do not include CR streaming in our simulations, only cosmic ray diffusion. Additionally, it is numerically much cheaper to handle CR transport by diffusion rather than streaming. Moreover, previous work suggest gamma-ray observations are in tension with the slower transport of CR streaming compared to more rapid CR transport by diffusion (\citealt[][]{chan2019cosmic,hopkins2021testing} with a higher than typical diffusion coefficient; cf. \citep[][]{strong2007cosmic}). Following this idea, we allow for CR streaming heating even though we do not include streaming itself. Using a diffusion model accelerates our simulations because the streaming model imposes more stringent time step constraints than those imposed by typical cosmic ray diffusivities.

\subsection{Simulation domain and initial conditions}

The elongated $2 \times 2 \times 40$ kpc three-dimensional box geometry represents an approximately parallel-plane section of a galactic disk. The maximum midplane height of $\pm 20$ kpc is chosen to ensure a realistic temperature distribution \citep{hill2012vertical}, while maintaining the parallel-plane approximation. The horizontal $x$ and $y$ coordinate sides have periodic boundary conditions, while the $z$ coordinate `top' and `bottom' have diode boundary conditions \hl{(zero-gradient boundary conditions which set inflowing velocities, when they arise, to zero instead)}, which avoids issues from gas fallback near the boundary \cite[e.g.][]{sur2016galaxy}. We do not include differential rotation effects as we model a patch at galactic center. We use static mesh refinement with a resolution of 31.25 pc for $|z| < 8 $ and  62.5 pc for $8 \ \textrm{kpc} < |z| < 20 \ \textrm{kpc}$. These higher resolution refinement regions are four times more extended than in \citet[][]{holguin2019role} in order to better capture the phase structure of the plasma farther out of the galaxy, simitar to the choices made in \citet[][]{peeples2019figuring}.

For the gas initial conditions, we use a vertical equilibrium density solution for a stratified, isothermal self-gravitating system \citep{spitzer1942dynamics, salem2014cosmic} projected to a stratified parallel-plane box \citep[e.g.][]{de2007generation, walch2015silcc,rathjen2021silcc, holguin2019role}. The unperturbed density distribution is as follows

\begin{equation}
\rho(z) = \begin{cases} \rho_{0} \hspace{0.1cm}\mathrm{sech}^{2}\left( \frac{z}{2 z_{0}} \right) \ \ \mbox{ $\rho(z) > \rho_{\textrm{halo}}$}  \\
                        \rho_{\textrm{halo}}                                       \mbox{\hspace{1.75 cm}otherwise,} \end{cases}
\label{spitzer}
\end{equation}

\noindent where $\rho_0$ is the initial midplane density and $z_0$ is the vertical scale height. We can define the initial disk surface gas density as $\Sigma_0 = \int_{-20 \textrm{kpc}}^{20 \textrm{kpc}} \ \rho (z) \textrm{d}z$. The density  $\rho_{\textrm{halo}} = 1.0 \times 10^{-28} \textrm{g/cm}^3$ is the initial density of the halo. 

In order to improve reproducibility, we reduce differences in initial star formation due to machine dependent effects by perturbing the density distribution as follows:

\begin{equation}
    \rho(z) = \rho(z) \left(  1+\delta_{\rho} \sum_{i=1}^{3} \rm{sin}(\pi x_{i}/L_{\delta})^2 /3 \right)
\end{equation}
where $x_i$ represents the spatial coordinates and $L_{\delta}$ is the perturbation wavelength in units of the highest resolution of $31.25$ pc. We choose the density distribution parameters (see Table \ref{sim_param_table}) such that $\Sigma_0 = 100 \ M_{\odot}/\rm{pc}^2$, which corresponds to the average surface density within a radius of 10 kpc in the isothermal self-gravitating solution. There are no stellar particles initially. After the bulk of star formation occurs (about 100 Myr) in the simulation, roughly 20 $M_{\odot}/\rm{pc}^2$ is converted to stellar particles. The gas initial temperature is $T=10^4$ K.  The initial magnetic field follows the density distribution as $B(z) \propto \rho(z)^{2/3}$ with the mid-plane value $B_0 \approx 3 \mu G$. The field is oriented along one of the horizontal directions.

From the initial conditions the gas radiatively cools and collapses to a smaller scale height distribution, until the densities become large enough (and the other criteria satisfied as described in Section \ref{sfeed}) for stellar particles to form and begin stellar feedback. We do not include any artificial pressure support, as resulting steady-state properties are insensitive to particular choices of pressure support \citep{kim2017three}. By 100 Myr, stellar feedback suppresses star formation and the simulations tends towards steady state.

\begin{table}
  \caption{Simulation model parameters}
  \begin{center}  
  	\leavevmode
    \begin{tabularx}{0.45\textwidth}{Xl}
    \hline\hline
  Halo                  \\ \hline 
   $M_{200}^{(1)}$ &   $10^{12}$M$_{\odot}$      \\
   $c^{(2)}$ &  12                   \\ \hline
   Disk                 \\ \hline
    $\rho_{o}^{(3)}$ & 5.64 $\times 10^{-23}$ g\ cm$^{-3}$ \\
    $z_{o}^{(4)}$ & 0.03 kpc \\
   $\Sigma_{o}^{(5)}$ & 100 M$_{\odot}$\ pc$^{-2}$ \\
	$T_{o}^{(6)}$    & 10$^{4}$ K \\
	$B_{o}^{(7)} = B_{o,x}$  & 3 $\mu$G \\ 
	$\delta_{\rho}$ & 0.1 \\ 
	$L_{\delta}$ & 8 $\delta$ x \\ \hline   
Star Formation  \\ \hline
$n_{\rm thresh}^{(8)}$ & 10 cm$^{-3}$ \\
$T_{\rm floor}^{(9)}$ & 300 K \\
$m_{*,{\rm min}}^{(10)}$ & 10$^{5}$ M$_{\odot}$ \\
$\epsilon_{\rm SF}^{(11)}$ & 0.05 \\ \hline
Stellar Feedback  \\ \hline
$f_{*}^{(12)}$ & 0.25 \\
$f_{cr}^{(13)}$ & 0.1 \\
$f_{*,\rm{mom}}^{14}$ & 5.0 \\
$\epsilon_{\rm SN}^{(15)}$ & 10$^{51}$erg/(M$_{\rm sf}$c$^{2}$) \\
$M_{\rm sf}^{(16)}$ & 100 M$_{\odot}$ \\
$T_{\rm{ion}}^{(17)}$ & 1.5e4 K \\ 
$f_{\rm{*}}^{(18)}$ & [0, 1] \\
$f_{\rm{*,\rm{EUV}}}^{(19)}$ & 0.1 \\ \hline
Cosmic Ray Feedback \\ \hline
$\kappa_{\rm{CR}}^{(20)}$ & [$10^{28}$, $10^{29}$] $\rm{cm}^{2}/s$  \\ 
$H_{\rm{cr}}^{(21)}$  & on/off  \\ \hline
\end{tabularx}
\label{sim_param_table}
\end{center}

\textbf{Notes.} From top to bottom the rows contain: (1) halo mass; (2) concentration parameter; (3) initial midplane density; (4) initial scale height of the gas disk; (5) initial gas surface density; (6) initial temperature; (7) initial magnetic field strength; (8) gas density threshold for star formation; (9) floor temperature; (10) minimum stellar population particle mass; (11) star formation efficiency; (12) fraction of stellar mass returned to the ISM; (13) fraction of supernova energy bestowed unto CRs; (14) SNe momentum feedback boost factor, (15) SN energy per rest mass energy of newly formed stars; (16) rest mass energy of newly formed stars per SN; (17) the transition temperature between neutral and ionized gas; (18) multiplicative factor scaling the net radiative heating, including both the stellar and metagalactic components; (19) escape fraction scaling the EUV component of radiative heating, modeling HII region photon escape; (20) CR diffusivity parallel to the magnetic field; (21) inclusion of CR streaming heating or not.

\end{table}

\section{Simulation results}
\label{simulation_results}

We run 6 simulations to 200 Myr, optimizing for computational expense and exploration of the CR and ionizing radiation parameter space. The simulations are summarized in Table \ref{sim_list}. In the short name for the simulations `k' is followed by either 28 or 29, referring to the log of $\kappa_{\rm{cr}}$, `Hrad' refers to the inclusion of radiative heating in both the FUV and EUV bands, and `Hcr' refers to the inclusion of cosmic ray streaming heating. All simulations include CRs. In all simulations that include radiative heating there is a reduction in the total EUV luminosity by a factor $f_{*,\rm{EUV}}=0.1$, which is a reasonable mean escape fraction for ionizing photons from HII regions, although in practice there is a lot of scatter \citep[e.g.][]{leitherer1995lyman,pellegrini2012optical, heckman2023global}. Due to our increase in the size of the highest resolution region, the added computational cost of $\kappa_{\rm{cr}} = 10^{29} \ \rm{cm}^{2}s^{-1}$ limits the number of simulations we can run at this higher CR diffusivity.

\begin{table*}
    
    \caption{Summary of the 6 simulations we conducted, spanning limiting cases in CR diffusivity, as well as radiative and CR streaming heating.}
    \label{sim_list}
  \begin{center}  
  	\leavevmode
    \begin{tabularx}{0.65\textwidth}{lllll}
    \hline\hline
    Sim Number              & Sim short name & Radiative heating & log $k_{\rm{cr}} \ [\rm{cm}^2 s^{-1}]$ & CR streaming heating     \\ \hline
    1 &    \rm{k28}        & No          & 28                & No  \\
    2 &      \rm{k28\_Hcr}          & No        & 28      & Yes \\
    3 &    \rm{k28\_Hrad}      & Yes       & 28      & No  \\
    4 &   \rm{k28\_Hrad\_Hcr}    & Yes       & 28       & Yes \\
    5 &     \rm{k29\_Hrad}  & Yes         & 29       & No  \\
    6 &     \rm{k29\_Hrad\_Hcr}   & Yes     & 29      & Yes \\
    \hline 
    \end{tabularx}
  \end{center}
\end{table*}

\begin{figure*}
    \centering
    \begin{subfigure}[t]{0.75\textwidth}
        \centering
        \includegraphics[width=0.9\textwidth]{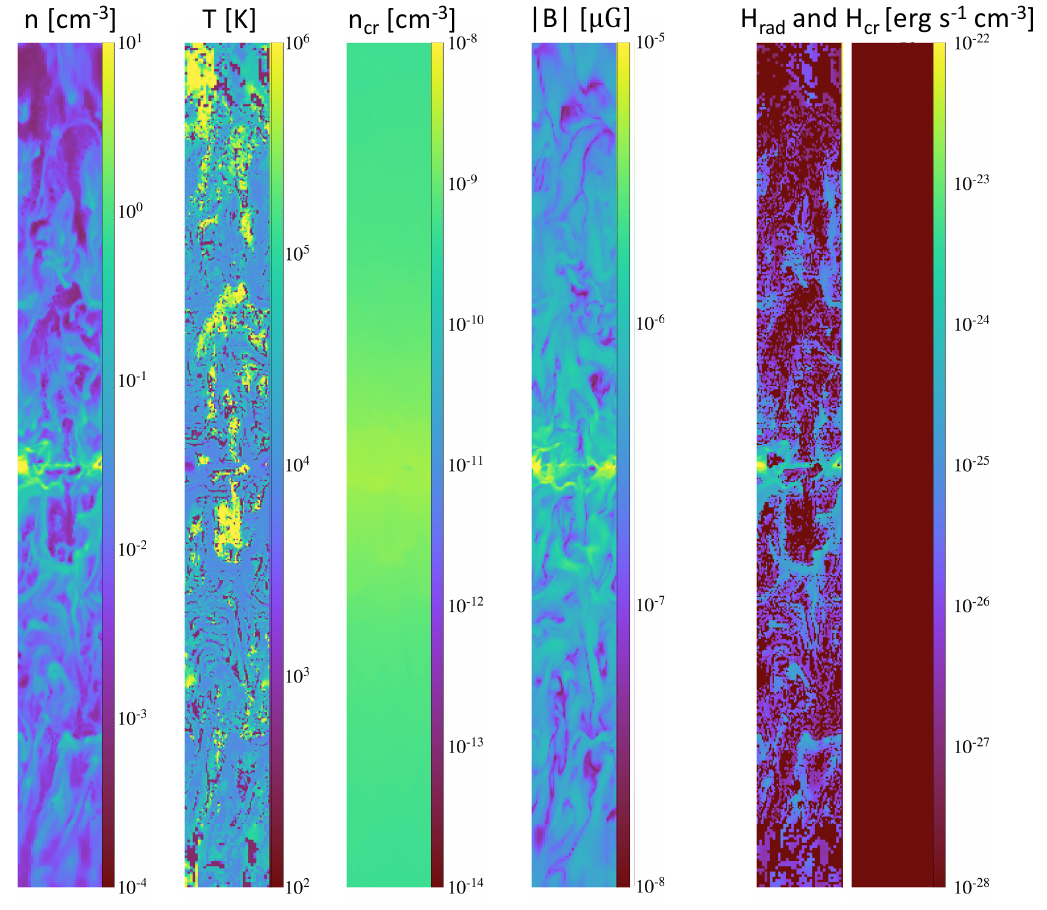}
        \caption{'k29\_Hrad': No CR streaming heating.}
        \label{slices_a}
    \end{subfigure}
    \hfill
    \begin{subfigure}[t]{0.75\textwidth}
        \centering
        \includegraphics[width=0.9\textwidth]{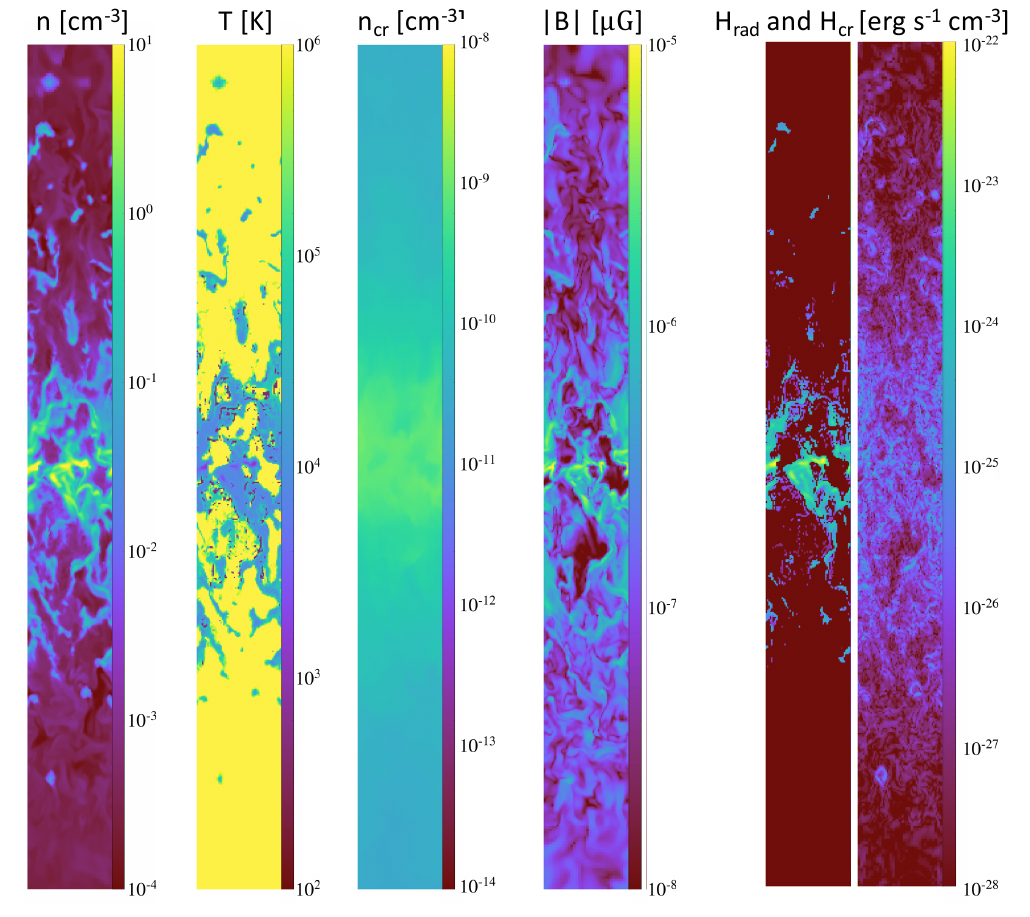}
        \caption{'k29\_Hrad\_Hcr':CR streaming heating included.}
        \label{slices b}
    \end{subfigure}
    
    \caption{Slice plots from simulations at 190 Myr, both with $\kappa_{\rm{cr}}=10^{29} \ \rm{cm}^2 s^{-1}$ and radiative heating. The slice are (left to right) number density $n_i$, temperature $T$, cosmic ray number density $n_{\rm{cr}}$, and heating rates (radiative $H_{\rm{rad}}$ and CR streaming heating rate $H_{\rm{cr}}$). The outflow in `k29\_Hrad' extends to the top of the domain at 20 kpc from the midplane. The inclusion of CR streaming heating losses reduces the CR population enough to reduce the extent of the outflow to roughly 10 kpc. }
    \label{slices}
\end{figure*}

For illustrative purposes, in Figure \ref{slices} we highlight both of the simulations with higher CR diffusivity,  `k29\_Hrad' and `k29\_Hrad\_Hcr', which produce extended outflows. The figure shows slices of a $x-z$ plane through the origin. The color in the slices denotes the value of $n_{\rm{i}}$, $T$, $n_{\rm{cr}}$, $H_{\rm{rad}}$, and $H_{\rm{cr}}$ along the plane. Radiative heating produces a volume filling $T \sim 10^{4}$ K medium around the galactic disk, unlike in our previous simulations in \citet[][]{holguin2019role}, where there is a significant amount of colder phases. The scale height of radiative heating is smaller compared to that of CR streaming heating, when it is enabled. The outflow is weaker with streaming heating present, although the CR distribution does extend far outside the galaxy for the weaker wind. This case suggests that the CRs are losing significant amounts of energy to the surrounding gas.

%\begin{figure*}
%    \centering
%    \includegraphics[width=0.95\textwidth]{grid_prop_rotate.png}
%    \caption{Summary of 6 simulations ran with profile plots of key quantities: $n$, $n_{\rm{cr}}$, total radiative heating rate $H$, and non-thermal pressure ratio $P_{\rm{cr}}/P_{\rm{th}}$, as well as number density $n$ and volume filling fraction $f_{\rm{vol}}$  for each gas phase used in Cloudy. The definitions of the phases are in Eq. \ref{phases}. The runs are plotted at 190 Myr. }
%    \label{sim_profile_grid}
%\end{figure*}

\begin{figure*}
    \centering
    \includegraphics[width=1\textwidth]{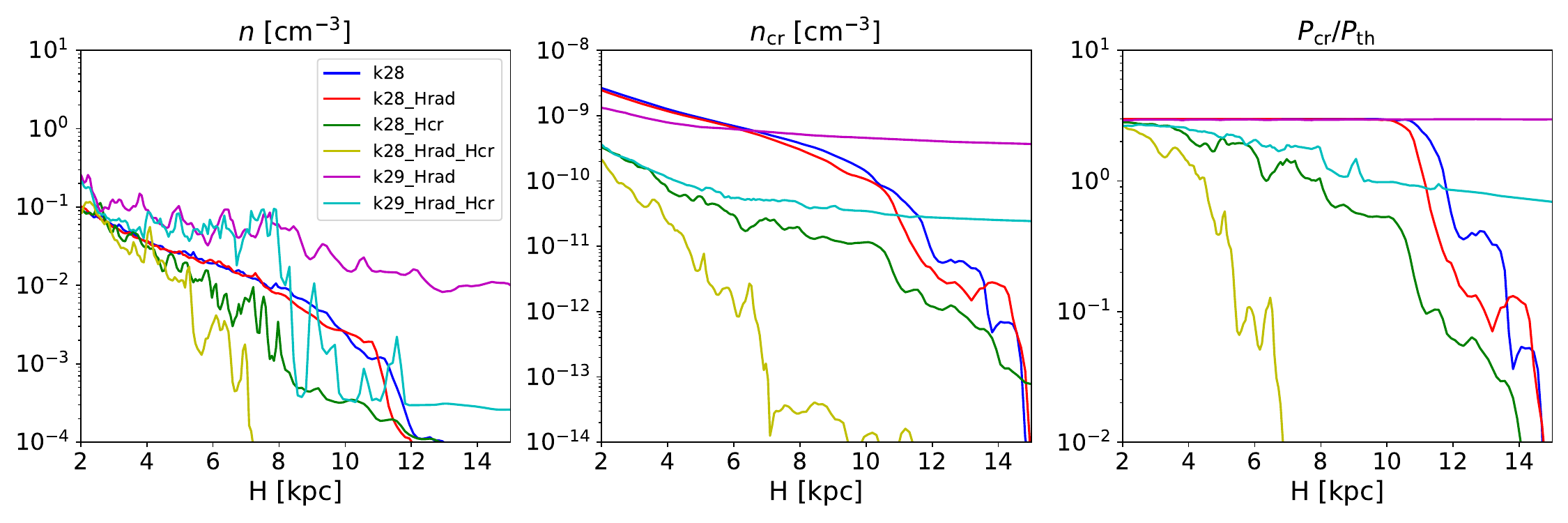}
    \caption{Profiles at 190 Myr from 6 simulations of gas and CR number densities, as well as ratio of CR to thermal pressure $P_{\rm{cr}}/P_{\rm{th}}$, vs. midplane height. The `k29' simulations with larger CR diffusivity, have a more extended outflow of gas compared to their `k28' counterparts. The inclusion of CR streaming heating suppressed the outflow. The lower  lower CR diffusivity simulation with CR streaming heating and radiative heating, resulted in the weakest outflows. }
    \label{density_pressure_overplot}
\end{figure*}

\begin{figure*}
    \centering
    \includegraphics[width=1.0\textwidth]{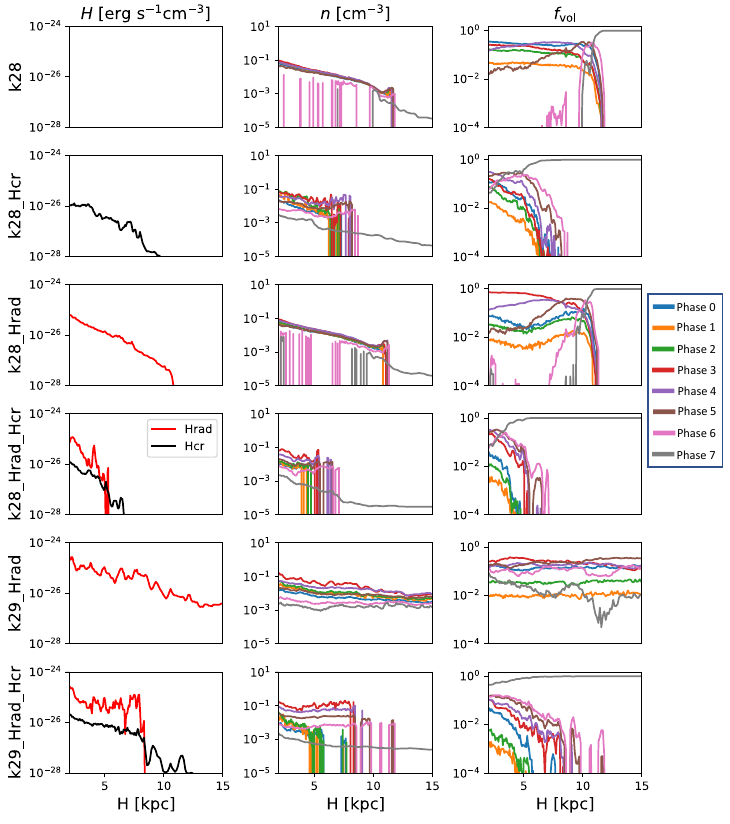}
    \caption{Profiles at 190 Myr from 6 simulations for theating rate from total radiative and CR streaming separately, $H$, (left), gas densities per thermal phase (middle), and volume filling fraction $f_{\rm{vol}}$  (right). The thermal phases, defined in Eq. \ref{phases}, correspond to the phases used in the Cloudy modeling. }
    \label{phase_profile_grid}
\end{figure*}

\begin{figure*}
    \centering
    \includegraphics[width=\textwidth]{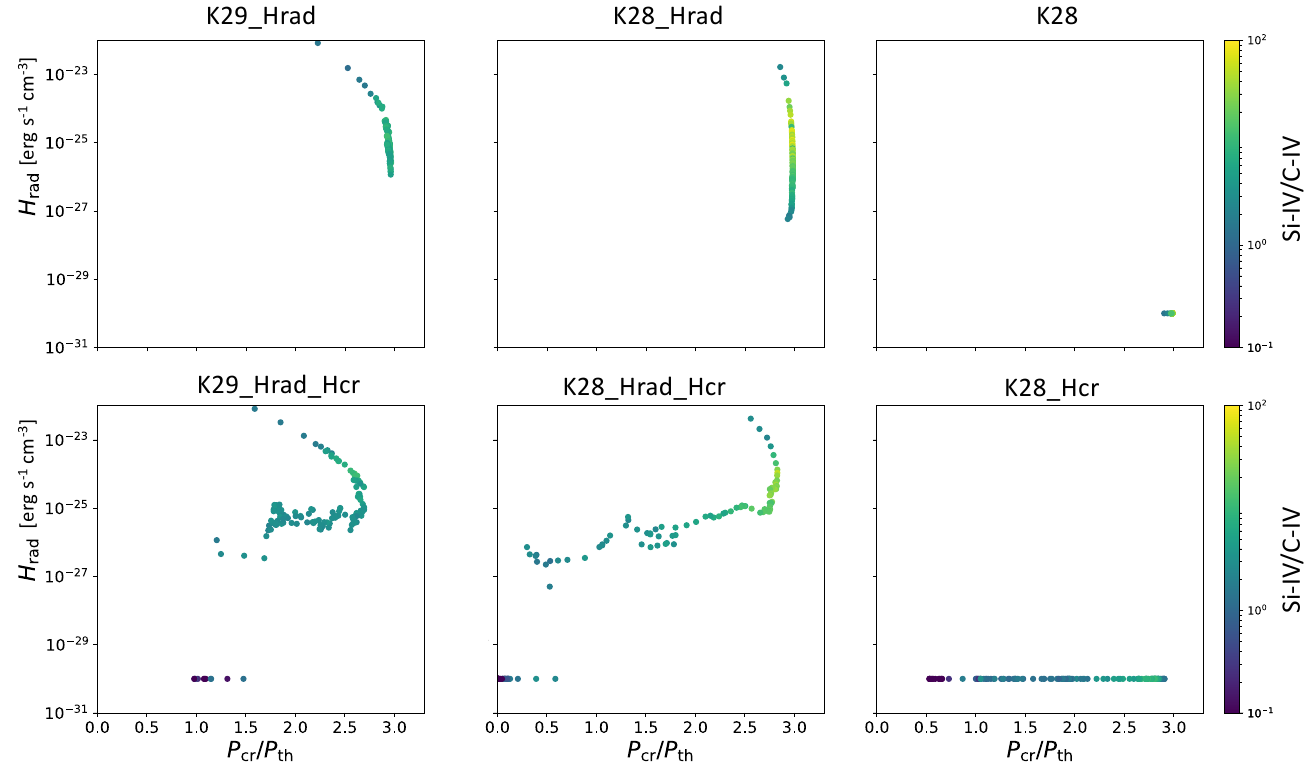}
    \caption{\hl{Scatter plot ($H_{\rm{rad}} [\rm{erg} \ \rm{s}^{-1} \rm{cm}^{-3}]$ vs. $P_{\rm{cr}}/P_{\rm{th}}$) of Si-IV/C-IV line ratio values for each slab in the simulation below 10 kpc, for each simulation type. The line ratios are discussed in} Section \ref{line_diagnostic_discussion}. \hl{For the simulations without radiative heating, K28 and K28\_Hcr, the data points were placed at an artificial floor of $H_{\rm{rad}} = 10^{-30} \rm{erg} \rm{s}^{-1} \rm{cm}^{-3}$ to aid visualization. Following the curves from high to low $H_{\rm{rad}}$ traces low to high midplane heights. The loss of CR pressure support is seen in simulations with CR collisionless heating.} }
    \label{siiv_civ_scatter}
\end{figure*}

\begin{figure*}
    \centering
    \includegraphics[width=0.5\textwidth]{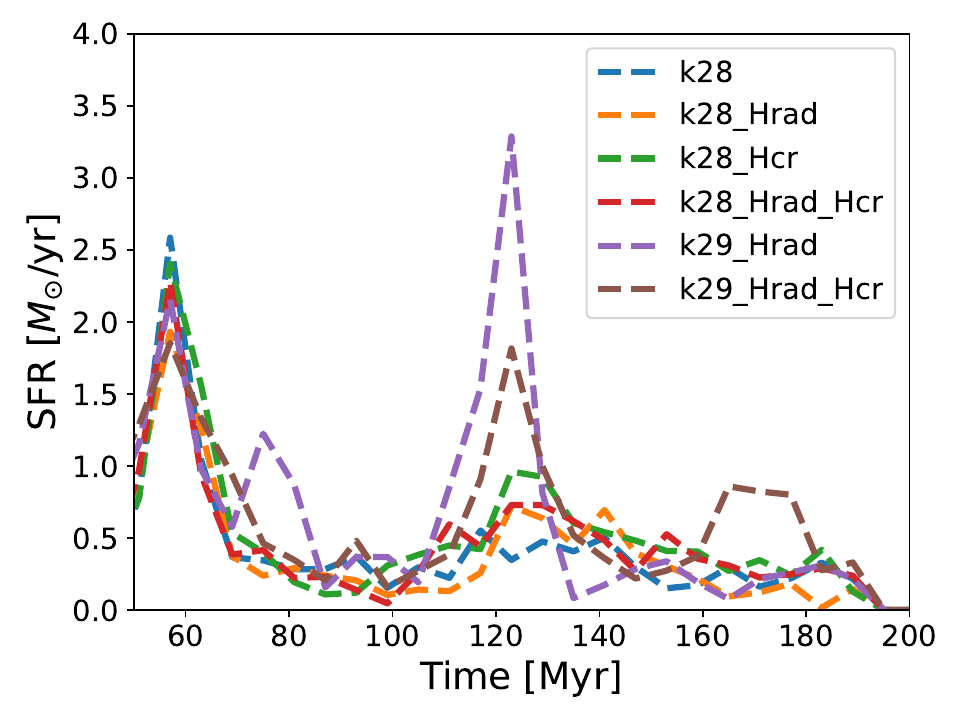}
    \caption{SFR of the 6 simulations we ran. There is an initial burst of star formation at 60 Myr. The simulations with higher $\kappa_{\rm{cr}}$ experience another, larger burst of star formation at 120 Myr, due to CRs leaving the dense disk more efficiently. This removal of pressure support triggers more gravitational collapse and star formation.}
    \label{sim_sfr_loading}
\end{figure*}

In Figures \ref{density_pressure_overplot}, \ref{phase_profile_grid}, \ref{siiv_civ_scatter} and \ref{sim_sfr_loading}, we compare the simulation results for all 6 of the simulations we conducted for key quantities. In Figure \ref{phase_profile_grid}, we refer to columns and rows relative to the label orientation, in that the columns show profile quantities, and the rows are of different simulations. The first column shows the heating rate from radiative and CR streaming. The last two columns show the number density and volume filling fractions of the phases used for post-processing in Eq. \ref{phases}. First, we describe simulations without CR streaming heating. Radiative heating has the effect of reducing the volume filling fraction of the cold neutral medium (phases 0, 1, and 2) while increasing the fraction of the warm neutral and ionized phases (phases 3 and 4). The hotter phases are not significantly affected, as radiative heating does not affect ionized plasma. The density profiles of the phases remain similar, as the phases are dominated by non-thermal pressure support instead of thermal pressure support, which is shown in the fourth column. The outflows for the lower $\kappa_{\rm{cr}}$ runs are weaker than in the higher $\kappa_{\rm{cr}}$ case, in agreement with previous works \citep[e.g.][]{salem2014cosmic,hopkins2020but,farcy2022radiation}. The outflows extend to roughly 10 kpc for lower CR diffusivity and 20 kpc for higher CR diffusivity. With higher CR diffusivity, the density profiles of the gas phases within 10 kpc are not so tightly overlapping, even though the non-thermal pressure support is dominant. The overlapping of phase densities does occur past 10 kpc. The dominant phases in the outflow are the hot phases (phases 6 and 7), unlike the lower $\kappa_{\rm{cr}}$ case. There is also a resurgence in star formation for higher $\kappa_{\rm{cr}}$, likely due to CRs more easily leaving the galactic disk and reducing pressure support after their initial injection by the first burst of star formation.

Simulations with CR streaming heating significantly reduce the spatial extent of outflows from the galaxy for both values of $\kappa_{\rm{cr}}$. The energy transfer to the gas may not have a large influence in its thermal energy for some parts of the cooling curve \citep[see][]{hopkins2020but}, but the energy loss can be significant to the CR population. In a CGM model with a fixed CR flux \citet[][]{huang2022launching} similarly found that CR streaming heating can be a dominant mechanism of energy transfer from the CRs. Two quantities point to the outflow weakness being caused by the reduction in CR energy density: $H_{\rm{cr}}/\Gamma_{\rm{cr}}$ (the CR energy loss rate relative to hadronic and Coulomb processes in Eq. \ref{cr_loss_eq_hcr}) and the loss timescale $\tau_{\rm{cr, st}} \sim e_{\rm{cr}}/ H_{\rm{cr}}$ (Eq. \ref{tau_eq}). CR collisional losses are proportional to gas density, so as the density decreases farther away from the galaxy, these losses are reduced. Setting the collisional and collisionless CR loss terms equal to each other yields a critical number density, 

\begin{equation}
\begin{split}
    & H_{\rm{cr}} \sim  \ \Gamma_{\rm{cr}} \\
    & v_{\rm{A}}\nabla P_{\rm{cr}} \sim  \Gamma_{\rm{cr}}  \\
        & v_{\rm{A}} \left( \frac{e_{\rm{cr}}}{L_{\rm{cr}}} \right)\sim  9.376 \times 10^{-28} \frac{\rm{erg}}{\rm{cm}^3 s} \left(\frac{n}{\rm{cm}^{-3}} \right) \left(\frac{e_{\rm{cr}}}{\rm{eV}\rm{cm}^3} \right) \\
        & 10^{3/2}\left( \frac{5.5 \times 10^{-28}}{9.376 \times 10^{-28}} \right)  B_{\mu G} \ \left( \frac{L_{\rm{cr}}}{\rm{kpc}} \right)^{-1} \sim  \left(\frac{n}{\rm{cm}^{-3}} \right)^{3/2}\\
     & \left(\frac{n}{\rm{cm}^{-3}} \right) \ \sim  \ 0.1 \ \ B_{\mu G}^{2/3} \  \left( \frac{L_{\rm{cr}}}{\rm{kpc}} \right)^{-2/3}
\end{split}
\label{cr_loss_eq_hcr}
\end{equation}
%    5.5 \times 10^{-28} \frac{\rm{erg}}{\rm{cm}^3 s} \left(\frac{e_{\rm{cr}}}{\rm{eV}\rm{cm}^3} \right) \left( \frac{B}{\mu \rm{G}} \right) \left( \frac{ L_{\rm{cr}}}{\rm{kpc}} \right) \left( \frac{n}{0.01 \rm{cm}^{-3}} \right)  \\
%    (x) \frac{e_{\rm{cr}}}{L_{\rm{cr}}} \sim & \Gamma_{\rm{cr}}  \\
%    n  \sim & 0.1 \ \rm{cm}^{-3} \ B_{\mu G} \ \frac{L_{\rm{cr}}}{\rm{kpc}}

where $L_{\rm{cr}}$ is the length scale of the global cosmic ray distribution and $B_{\mu G}$ is the magnetic field strength in $\mu G$. Below this density, collisionless losses dominate over collisional losses. For reasonable parameters in the low density medium above the galaxy, CR streaming losses should be the dominant loss mechanism below $n \sim 0.1 \ \rm{cm}^{-3}$. We also calculate the timescale of CR energy loss via CR streaming heating,

\begin{equation}
    \begin{split}
        \tau_{\rm{cr, st}} = & \ e_{\rm{cr}}/ H_{\rm{cr}} \\
                               = & \ e_{\rm{cr}} / | v_{\rm{A}}\cdot \nabla P_{\rm{cr}}| \\
                               \sim & \ 100 \ \rm{Myr} \ \left( \frac{L_{\rm{cr}}}{\rm{kpc}}\right) \left(\frac{n}{\rm{cm^{-3}}}\right)^{1/2} B_{\mu G}^{-1}, 
    \end{split}
    \label{tau_eq}
\end{equation}
which indicates that for reasonable parameters, the CR energy loss timescale due to collisionless processes is of order the total simulation time of 200 Myr, or shorter in lower density regions. The column for $P_{\rm{cr}}/P_{\rm{th}}$ in Figure \ref{phase_profile_grid} shows that in the simulations with CR streaming heating, CRs retain pressure dominance in the disk (within 3 kpc) but farther out of the galaxy, thermal pressure dominates, as expected by our analysis above. In the heating rate column, we note that CR streaming heating does not match the radiative heating rate until roughly 7-10 kpc in simulations with both heating mechanisms included. Although, it is likely we could be overestimating the radiative heating rate farther away from the galaxy since a non-attenuated parallel-plane radiation field does not decay with height. In global simulations, the radiative heating rate could fall faster with height above the disk (as the stellar field decays), allowing other heating mechanisms like CR streaming heating to dominate for a larger range in height.

\hl{Figure} \ref{siiv_civ_scatter} \hl{ provides another perspective of the radiative heating and pressure state of the galaxy. This figure highlights the varied combinations of radiative heating and pressure support states of the gas depending on the feedback mechanisms included. In the simulations without collisionless CR losses, the CR-dominated pressure remains similar across all radiative heating values. Comparing the  `k28' and `k29' simulations, we see a loss of CR pressure support in the midplane with faster CR diffusivity, as the CRs more quickly diffuse out of the disk. The slabs approach the CR-dominated values of pressure at roughly 2 kpc in height. In the `Hcr' cases with collisionless CR losses, the CR pressure support declines with height in between 2-5 kpc. Interestingly, the `k28\_Hrad\_Hcr' case contains gas that exists at all levels of CR pressure support between the highest level and thermally dominated.} \hl{Figure} \ref{siiv_civ_scatter_hradcr} \hl{shows the same plot, except with the $H_{\rm{cr}}$ contribution to heating included. The main difference in this plot is that the heating rate for data points farther away from the disk, with lower $P_{\rm{cr}}$ and $H_{\rm{rad}}$, is larger so that the curve decreases smoothly with the heating rate. Further discussion of predicted plasma diagnostics associated with these plots occurs in later sections. }

\section{Post-processing framework with Cloudy}
\label{postprocessing_section}

\subsection{Division of simulation domain and Cloudy model setup}
\label{domain_setup}
In order to produce synthetic plasma diagnostics from our simulations, we use the spectral synthesis code, Cloudy \citep[][]{ferland2017}. A typical Cloudy parallel-plane model takes inputs of hydrogen density, plasma temperature,  metalicity, and incoming spectra (radio to x-ray), resulting in outputs of species ionization state and outgoing spectra. We use the MHD simulations to inform various Cloudy models. Cloudy is able to calculate the plasma temperature. However, this calculation does not include the various feedback mechanisms present in the three dimensional MHD simulation, so we let the model temperature be informed by the simulation. We calculate the galactic midplane spectra $J_{\nu,*}$ by assigning spectra to each simulation stellar population particle based on particle age and mass using Starburst99. 

\begin{figure*}
    \centering
    \includegraphics[width=0.8\textwidth]{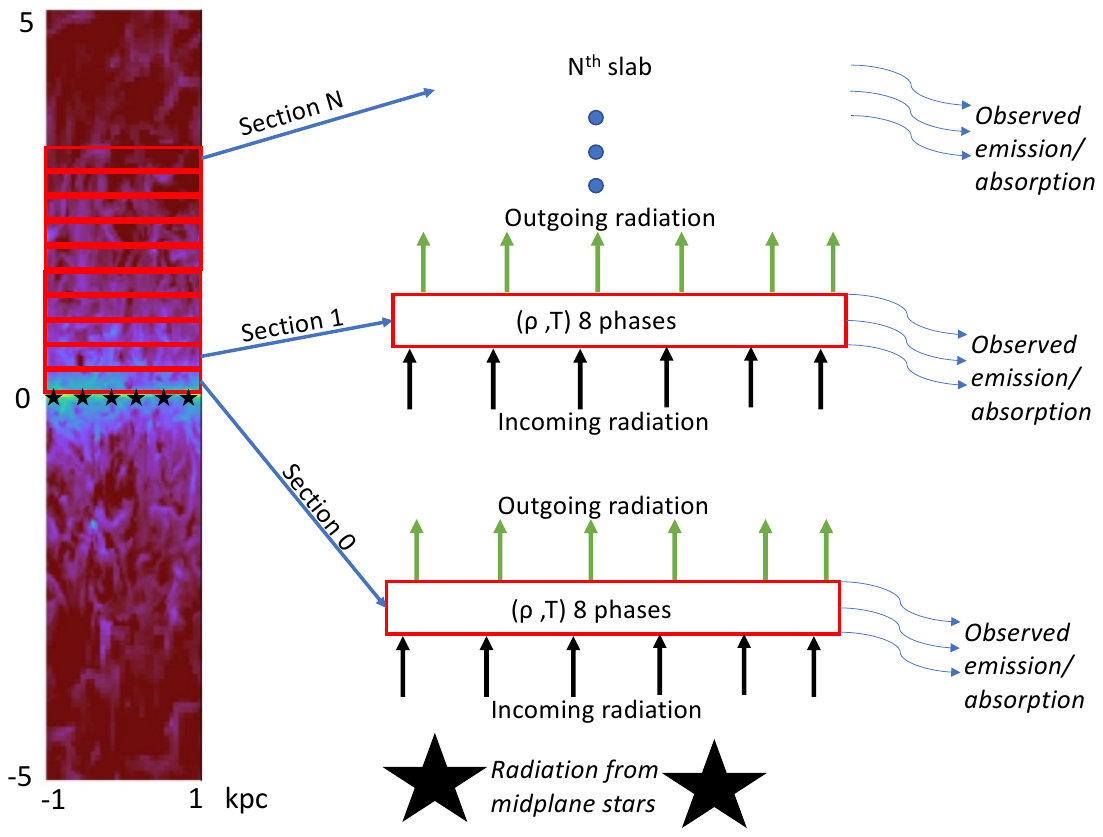}
    \caption{Diagram of MHD simulation and Cloudy post-processing framework. A slice ($2 \times 2 \times 5$ kpc) of an example simulation is shown divided up into red parallel plane sections, the properties of which are used to inform Cloudy models. Each section holds eight temperature phases, from Eq. \ref{phases}, each with mean density $\rho$, temperature $T$ and volume filling factor $f_{\rm{vol}}$. The midplane spectra from the stellar population particles is transmitted outward through the sections. We calculate the ionization state of H, C, N, O, and Si ions in each section. }
    \label{cloudyframework}
\end{figure*}

Figure \ref{cloudyframework} shows a diagram that illustrates the post-processing framework. We divide the simulation domain into parallel-plane sections of two resolution cells in height. Within each section $i$, we calculate the mean and standard deviation of $n$ and $T$ of each gas phase. We define eight gas phases in each section $i$ from the temperature floor to hot coronal gas as follows,
\begin{equation}
\begin{split}
\rm{Phase\ 0}:& \ T < 500 \rm{K} \\
\rm{Phase\ 1}:& \ 500 \rm{K} < T < 2000 \rm{K} \\
\rm{Phase\ 2}:& \ 2000 \rm{K} < T < 8000 \rm{K} \\
\rm{Phase\ 3}:& \ 8000 \rm{K} < T < 1.5 \times 10^{4} \rm{K} \\
\rm{Phase\ 4}:& \ 1.5 \times 10^{4} \rm{K} < T < 3 \times 10^{4} \rm{K} \\
\rm{Phase\ 5}:& \ 3 \times 10^{4} \rm{K} < T < 10^{5} \rm{K} \\
\rm{Phase\ 6}:& \ 10^{5} \rm{K} < T < 10^{6} \rm{K} \\
\rm{Phase\ 7}:& \ 10^{6} \rm{K} < T
\end{split}
\label{phases}
\end{equation}
which include rough matches to the traditional ISM phases (phase 3: warm, neutral medium; phase 4: warm, ionized medium) \citep[][]{draine2010physics}. Each phase $j$ has mean density $\langle \rho_{i,j} \rangle$, temperature $\langle T_{i,j} \rangle$, and associated standard deviations $\sigma_{\rho, i}$ and $\sigma_{T,i}$ respectively, as well as the filling fraction $f_{\rm{vol;i,j}}$ of each phase in that section. A Cloudy model is assigned density $\rho_{i,j}^{\rm{C}}$ and temperature $T_{i,j}^{C}$ by drawing from a Gaussian distribution defined by the mean and standard deviation of density and temperature of each phase, with values limited to be within one standard deviation. We draw properties 10 times in order to produce 10 separate profiles of density and temperature in the domain. These separate profiles allow us to produce intervals in the synthetic line ratio profiles. The incoming spectra for each model is the midplane spectra $J_{\nu; *}$ if $i=0$ (lowest height section) or $J_{\nu; i-1}$ (the outgoing spectra from section below the current one) otherwise. The output spectra of the Cloudy model is $J_{\nu; i,j}$. The total outgoing spectra $J_{\nu, i}$ for section $i$ is the sum of the eight spectra in each phase $J_{\nu; i,j}$ weighted by the phase filling fraction $f_{\rm{vol;i,j}}$, which becomes the input spectra to the Cloudy models in the next section $i+1$. Through this framework, we obtain estimates of the ionization state of numerous species and are then able to generate plasma diagnostics that are spatially (with midplane height) and temporally dependent.

The parameter $f_{\rm{*}}$ is a binary value that activates or suppresses the presence of radiative heating. Due to limited resolution, we do not resolve the giant star forming complexes where massive stars form. We multiply the ionizing part of the stellar spectrum by an escape fraction $f_{\rm{*,\rm{EUV}}}$ to account for the absorption of hydrogen ionizing photons after emission from stars and out into the ISM. We also have $f_{\rm{bkg,\rm{EUV}}}$, the fraction of the metagalactic background field present. $f_{\rm{bkg,\rm{EUV}}}$ is a binary on/off option. Since the post-processing models the outward and not inward propagation of radiation, the metagalactic background field must be included in each Cloudy model as a binary on/off option. The metagalactic field is added to a Cloudy model by specifying the desired field name. We use `HM12' \citep[][]{haardt2012radiative} for our work. In order to avoid adding the previous slab attenuated background field every time we transmit a spectra to a new  domain section, we add the `no isotropic' option, which automatically removes the attenuated component of the isotropic spectra sources. Overall, we have consistency between the the choices in post-processing parameters and simulation parameters (e.g., same $f_{\rm{*}}$, $f_{\rm{*,\rm{EUV}}}$ and $f_{\rm{bkg,\rm{EUV}}}$  in simulation and post-processing). 

\subsection{Observational absorption-line diagnostics}
\label{line_diagnostic_discussion}

\begin{figure*}
    \centering
    \includegraphics[width=\textwidth]{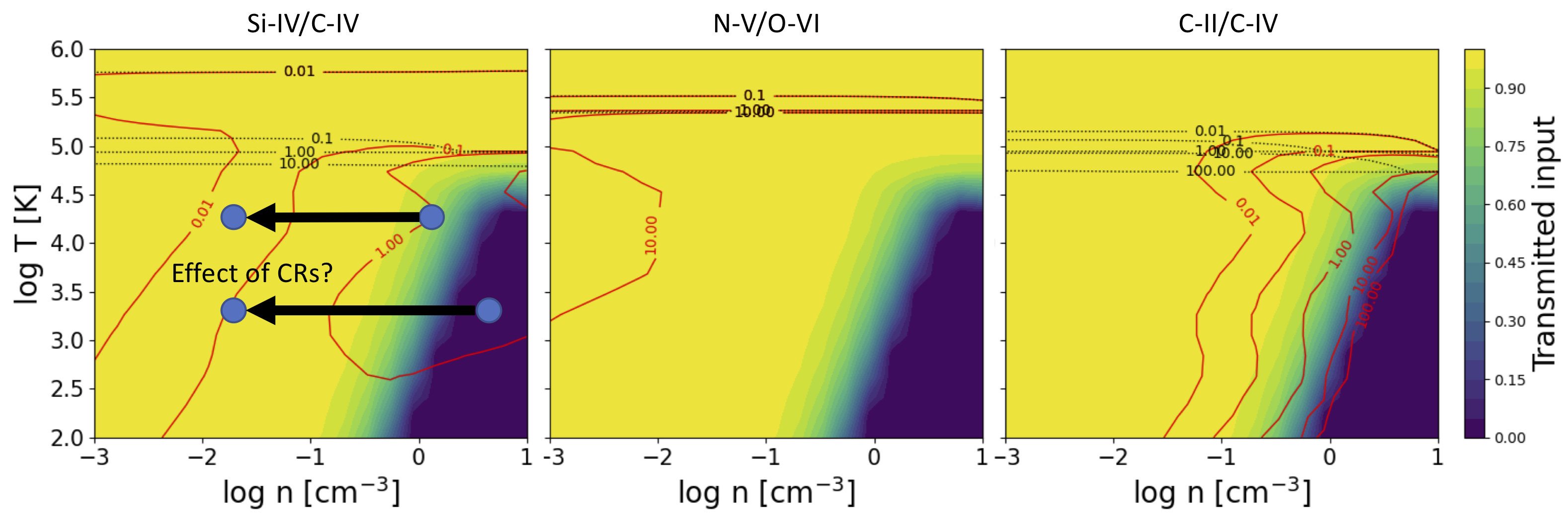}
    \caption{Line ratio diagnostics from Cloudy \hl{models} in density and temperature space for a 50 pc slab ($\sim 2$ high resolution cell widths \hl{in} our MHD simulations). \hl{These models are not informed by MHD simulations.} The black dotted contours show the line ratio values assuming collisional ionization, while the red contours show the values with the addition of photoionization from a constant SFR stellar population spectrum (SFR = $1 \ \rm{M}_{\odot}/\rm{yr}$). The background color shows the transmission fraction through the slab for reference. The addition of the stellar field significantly changes the line ratio contours at $T < 10^5$ K, except for N-V/O-VI. We sketch the potential effect of non-thermal pressure support on the Si-IV/C-IV plot. The right pair of blue dots are placed roughly where the traditional warm, ionized medium and cold, neutral phases exist based on thermal equilibrium. When the dominant pressure source is non-thermal pressure, multiple phases can exist at the same density \citep[e.g.][]{ji2020properties}. When these phases are shifted to lower densities, the line ratio values are closer to Milky Way values. In a realistic system the dominant pressure support will vary between thermal and non-thermal, so the analysis here provides limiting cases. }
    \label{theoretical_line_ratio}
\end{figure*}

Absorption-line ratios are common tracers of the gas kinematic and plasma state. Their values can distinguish between various multiphase non-equilibrium models (e.g. radiative turbulent mixing layers \citep[][]{ji2019simulations}, conductive interfaces \citep[][]{borkowski1990radiative, marcolini2005dynamics, bruggen2016launching}. Line ratio diagnostics are convenient to use because our parallel-plane domain represents only a piece of the galaxy, preventing an accurate prediction in the total ion column densities. We focus on three relatively common diagnostic absorption-line ratios: two intermediate (ionization energy > 30 eV) ion ratios, Si-IV/C-IV and  N-V/O-VI, as well as one low-to-intermediate ion ratio, C-II/C-IV \citep[][]{tumlinson2017}. These diagnostics span temperature ranges from the warm, neutral phases (phases 3,4) to just below the hot phase (phase 5). 

In order to explore the temperature and density parameter space that the simulation post-processing will effectively draw from, we run a grid of 20 by 20 data points of Cloudy models in $n_{\rm{i}}=[10^{-3},10^{1}]\ \rm{cm}^{-3}$ and $T=[10^{2},10^{6}]$ K, evenly spaced in log space, similar to common analyses done in observationally-focused work \citep[e.g.,][]{fox2005multiphase}. \hl{This analysis allows us to gain intuition for the potential interaction between radiative and CR feedback mechanism.} The Cloudy models all assume the same domain thickness of 50 pc, roughly the size of two of the highest resolution elements in the MHD simulations. \hl{The models in this figure are are not informed by the simulation results. The density and temperature values in this analysis do not represent upper or lower limits for the Cloudy models in subsequent sections that are informed by simulation results.} The input radiation field is that of the SFR = 1 $M_{\odot}$/yr stellar population spectra from Starburst99, multiplied by an escape fraction $f_{\rm{*,\rm{EUV}}}=0.1$. The results are shown in Figure \ref{theoretical_line_ratio} for Si-IV/C-IV, N-V/O-VI, and C-II/C-IV from left to right. The black line contours correspond to the line ratio values without any external radiation field. The red line contours show the line ratio values with the stellar field included. The background filled color corresponds to the transmission fraction of the incoming spectra through the slab. This color is plotted to illustrate the optical thickness of the medium to the spectra.

The relatively constant value of Si-IV/C-IV $\sim 0.2$ observed in the Milky Way \citep[][]{savage2009extension, werk2019nature} is a useful guide for examining the line ratio parameter space. In collisionally ionized gas, the possible gas phases for that line ratio are restricted to a narrow temperature range around $T \sim 10^{5}$ K: the value of the line ratio is also quite sensitive to the temperature around $T \sim 10^{5}$ K, with two orders of magnitude changes in value. With a SFR = 1 $\rm{M}_{\odot}/yr$ stellar field added, the line ratio contours are quite different at $T < 10^{5.5}$ K. Lower temperature models in the range $T=[10^{3}, 10^{4}]$ K are able to produce Si-IV/C-IV $\sim 0.2$, but only at low densities of $n \sim 10^{-1.5}\ \rm{cm}^{-3}$. Under standard ISM thermal pressure balance \citep[][]{mckee1977theory,draine2010physics}, these phases of gas are required to be much denser (at lease an order of magnitude) to remain in pressure equilibrium with the hot gas phase. By providing a non-thermal pressure source, CRs can allow multiple phases to exist at the same density \citep[see][]{ji2020properties}, in particular, by allowing more tenuous cold phases. The overplotted blue circles roughly mark where (right) thermal and (left) non-thermal gas phases exist. Whereas in the thermally supported case, lower temperature phases cannot reproduce observed line ratio values even with an external radiation field, with CRs, a variety of phases from $T = 10^3$ K to $T=10^5$ K  could reproduce the observed line ratio values.

The parameter space for C-II/C-IV is similar to that of Si-IV/C-IV. The addition of a stellar radiation field allows phases with $T < 10^{5}$ K to produce lower line ratio values. With non-thermal pressure support, these tenuous, cold phases can plausibly exist and contribute to the observed signal. The N-V/O-VI line ratio is more resistant to change by the external stellar field, as expected by the higher ionization energy of the ions. Typical observed values of the line ratio are in the range $[10^{-1.5}, 10^{0.5}]$ \cite[][]{wakker2012characterizing}, which are found in a small part of the parameter space in Figure \ref{theoretical_line_ratio}, favoring a collisionally ionized medium. Some line values of roughly $10^1$ can occur at $\sim 10^{4}$ K with a stellar radiation field present, but only in very tenuous gas in this phase with $n < 10^{-2} \ \rm{cm}^{-3}$. We additionally tested (not shown here) the effect of only the metagalatic background field and found that it produces effects in between that of the collisional ionization and stellar field cases. The metagalactic background field itself is not sufficient to change the parameter space significantly.

Besides pressure support, CR streaming heating could have an impact on the gas thermal state. Radiative heating does not occur in a significantly ionized gas, so the radiation field does not heat this gas (although radiative effects on the line emission/absorption still can occur). \citet[][]{wiener2013cosmicheating} suggested that because the CR scale height extends well outside the galactic disk and collisionless CR streaming heating affects the ionized medium, CRs could provide a plausible heating mechanism to explain anomalously high electron densities outside the disk \cite[][]{reynolds1999evidence}. Regardless of the pressure equilibrium (thermal or non-thermal), additional energy injection into the ISM/CGM plasma reduces the value of the Si-IV/C-IV and C-II/C-IV line ratios in Figure \ref{theoretical_line_ratio}.

\section{Post-processing results}
\label{results}

\begin{figure*}
    \centering
    \includegraphics[width=0.9\textwidth]{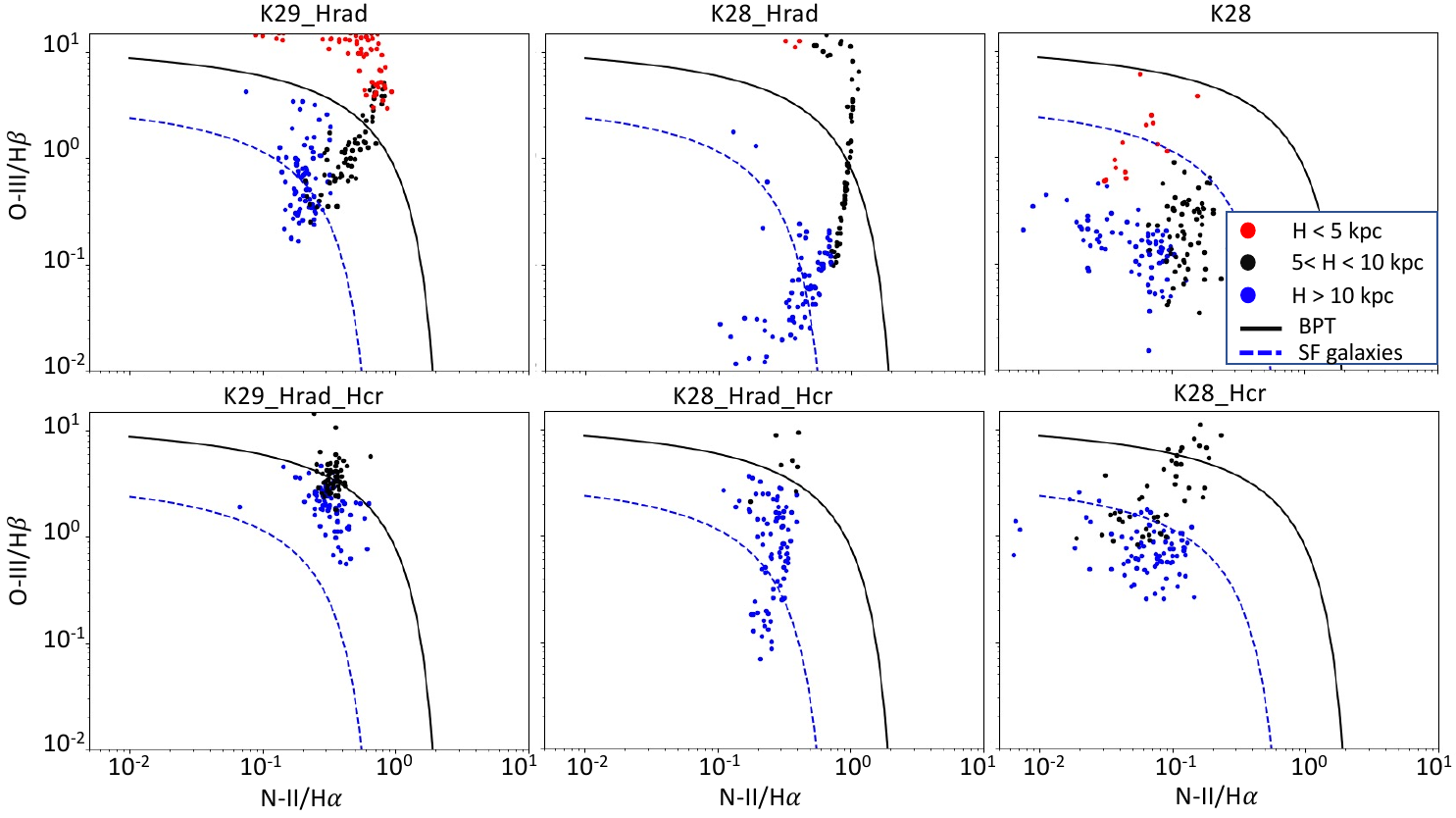}
    \caption{BPT diagram of 6 simulations. The solid line denotes the general line between star forming and active galaxies fitted from the \citet[][]{kewley2001theoretical} fit, while the dashed line denotes the rough trend for star forming galaxies. The simulations are shown at 190 Myr (same time shown in Figure \ref{slices}). Generally, the regions closer to the galactic disk (< 5 kpc) in all of the simulations were roughly on the star forming sequence, except for `k28', which did not include radiative or CR streaming heating. }
    \label{grid_bpt}
\end{figure*}

\begin{figure*}
    \centering
    \includegraphics[width=0.9\textwidth]{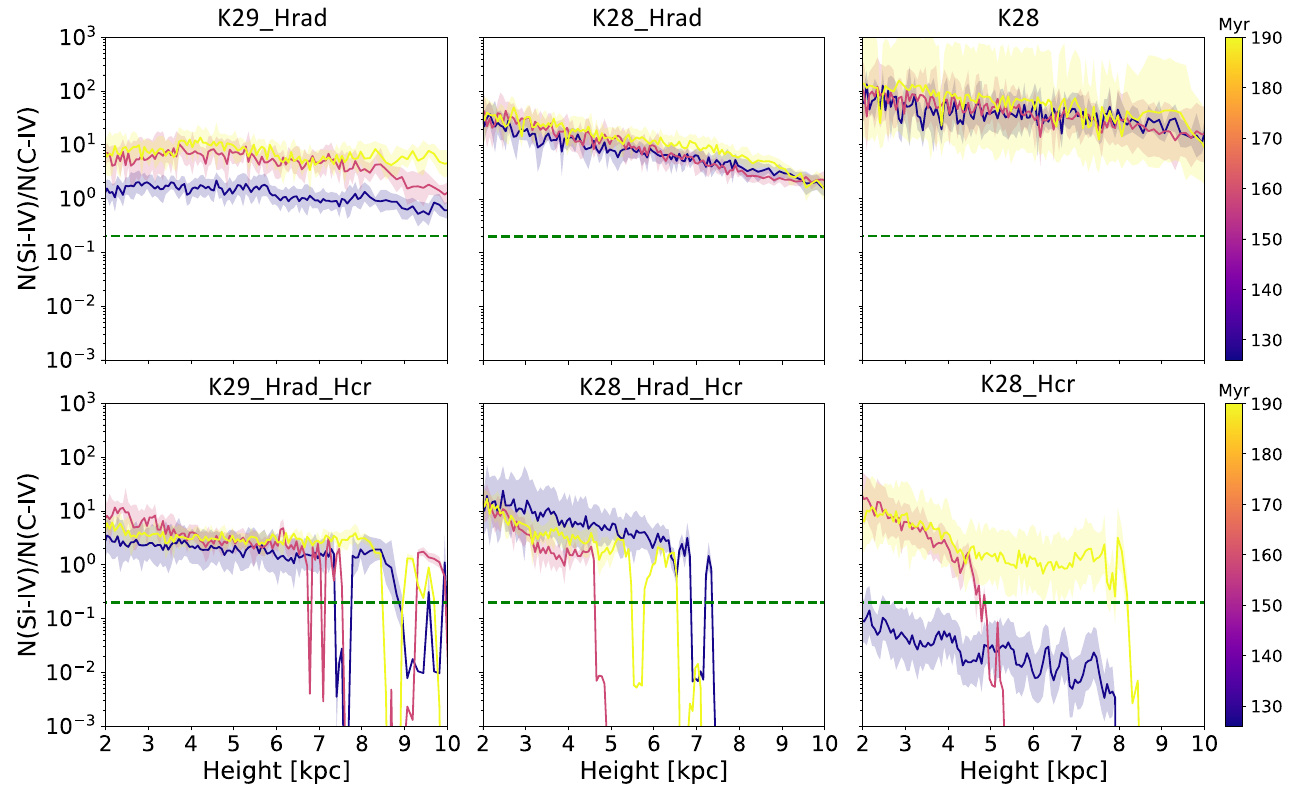}
    \caption{Line ratio profiles of Si-IV/C-IV for 6 simulations, colored by simulation time. The dashed, green line denotes the observed value in the Milky Way \citep[][]{werk2019nature}. For high CR diffusivity, `k29\_Hrad' (not including CR streaming heating) had the best match to Milky Way value. For low CR diffusivity, the addition of CR streaming heating reduced the line ratio values towards observed values, but the gas outflow was weak. }
    \label{siiv_civ}
\end{figure*}

\begin{figure*}
    \centering
    \includegraphics[width=0.9\textwidth]{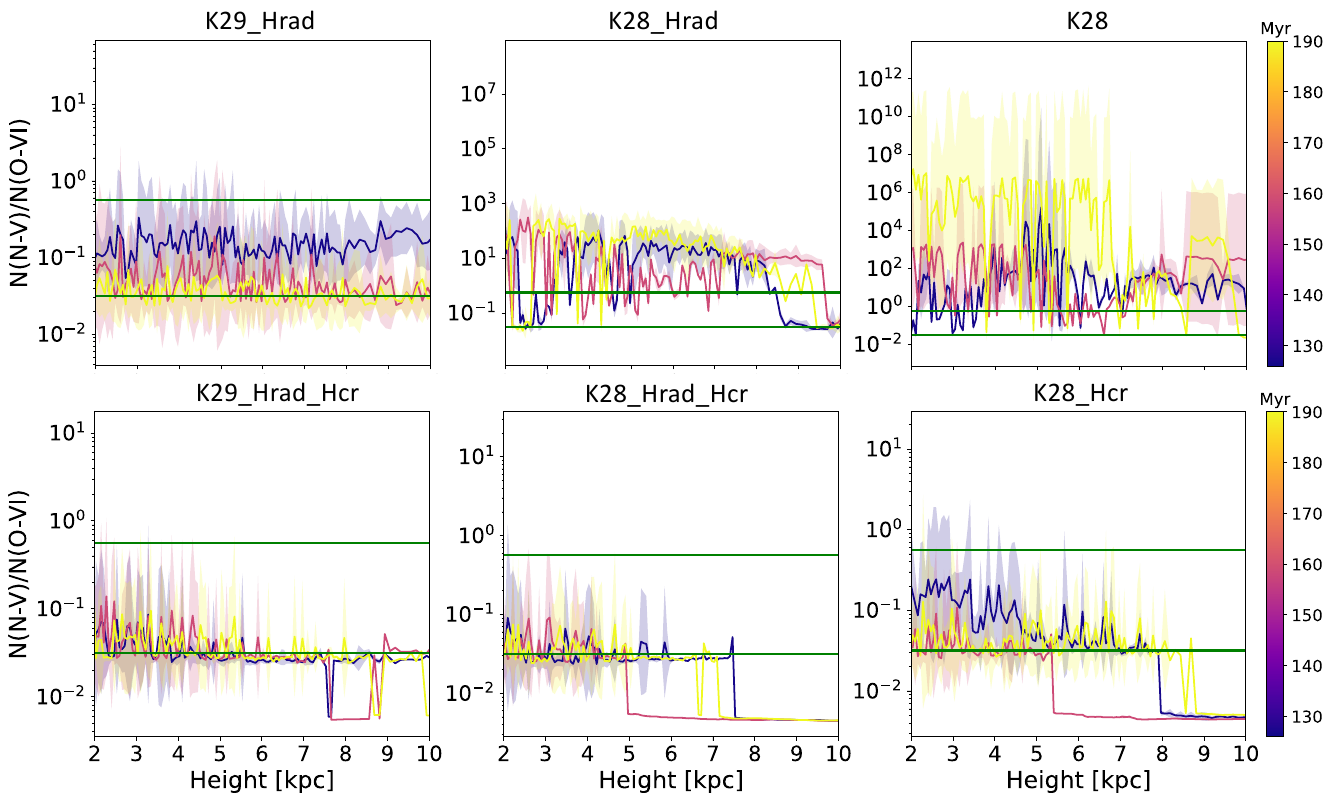}
    \caption{Line ratio profiles of N-V/O-VI for 6 simulations, colored by simulation time. The solid green lines denote the rough range of observed values compiled in \citet[][]{wakker2012characterizing} from many sources. Both simulations with high CR diffusivity have line ratio values within the observed range, while for low CR diffusivity, only the simultions with CR streaming heating have line ratio values consistently within the range.}
    \label{nv_ovi}
\end{figure*}

\begin{figure*}
    \centering
    \includegraphics[width=0.9\textwidth]{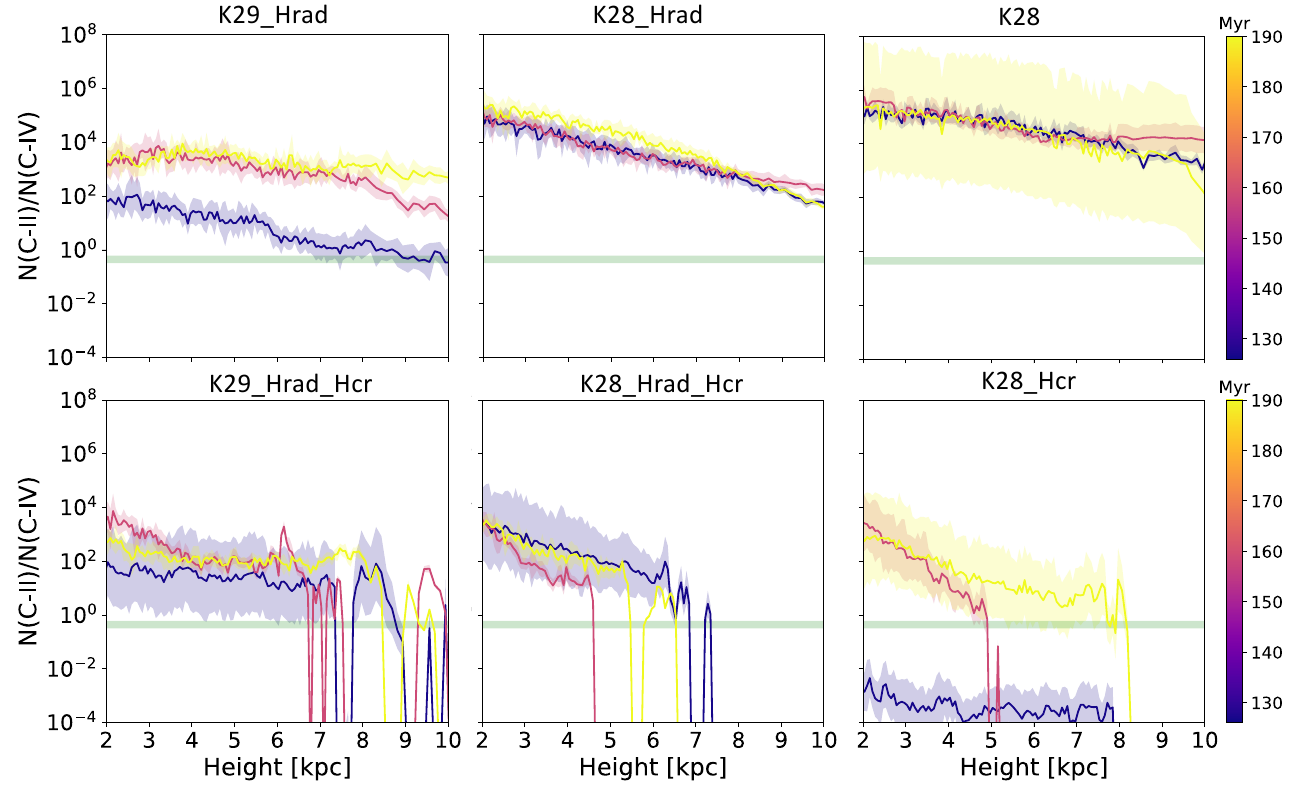}
    \caption{Line ratio profiles of C-II/C-IV for 6 simulations, colored by simulation time. The green range denotes the mean value $\pm$ standard deviation of the ratio from data in \citet[][]{fox2005multiphase}. This value is meant to provide a rough guide for typical values around the Milky Way halo. The results were similar to that of Si-IV/C-IV in Figure \ref{siiv_civ}, where the simulations closer to observed values were `k29\_Hrad' and `k28\_Hcr' at high and low CR diffusivity respectively.}
    \label{cii_civ}
\end{figure*}

%\begin{figure*}
%    \centering
%    \includegraphics[width=0.9\textwidth]{siiv_civ_hrad_pres_scatter.pdf}
%    \caption{\hl{Scatter plot ($H_{\rm{rad}} [\rm{erg} \ \rm{s}^{-1} \rm{cm}^{-3}]$ vs. $P_{\rm{cr}}/P_{\rm{th}}$) of Si-IV/C-IV line ratio values for each slab in the simulation below 10 kpc, for each simulation type. For the simulations without radiative heating, K28 and K28\_Hcr, the data points were placed at an artificial floor of $H_{\rm{rad}} = 10^{-30} \rm{erg} \rm{s}^{-1} \rm{cm}^{-3}$ to aid visualization. Following the curves from high to low $H_{\rm{rad}}$ traces low to high midplane heights. The loss of CR pressure support is seen in simulations with CR collisionless heating.} }
%    \label{siiv_civ_scatter}
%\end{figure*}

Section \ref{line_diagnostic_discussion} demonstrates influence of the gas thermal state and local radiative field on values of C-II/C-IV, Si-IV/C-IV and N-V/O-VI diagnostics. In a realistic system, many physical processes are acting in a complex, interdependent fashion, making the inverse problem of determining the gas state from spectral observations particularly challenging. We further explore this parameter space using the three-dimensional galaxy simulations from Section \ref{simulation_results}, which capture key parts of the nonlinear galactic system. The simulated CGM diagnostics can help disintagle the importance of various  physical processes through testable predictions. We post-process our simulations using the method described in Section \ref{domain_setup}.

\subsection{BPT diagram}

Figure \ref{grid_bpt} shows where each simulation falls on the BPT diagram \citep[][]{baldwin1981classification}, color coded for different distances from the midplane. Each axis on the plot is a line ratio in emission. The BPT diagram is commonly used to divide observed galaxies into those whose HII regions are primarily ionized by stars and those primarily ionized by an active-galactic nuclei (AGN). In other words, the dividing line on the BPT diagram separates galaxies into those ionized by a softer or harder spectrum respectively \citep[][]{veilleux1987spectral}. 

The BPT diagrams are useful in that they help summarize the properties of the ionizing radiation field, as well as ISM plasma conditions and metalicity \citep[][]{kewley2013theoretical}. The data points from `k28' are below the typical star forming sequence, as expected by the lack of stellar radiation present in the run. The addition of only CR streaming heating in `k28\_Hcr' moves the data points vertically in O-III/H$\beta$, overlapping star forming sequence. The addition of stellar radiation for low CR transport speed tends to move the near-disk data points horizontally and narrows values in N-II/H$\alpha$. For `k28\_Hrad\_Hcr' that includes both radiative and CR streaming heating, we see the combined effects on the BPT diagram from adding each heating mechanism separately.  The  near disk data points in the higher CR transport simulation (`k29\_Hrad') with only radiative heating are roughly around the star forming sequence, at a similar N-II/H$\alpha$ and higher O-III/H$\beta$ compared to the lower CR transport version. The addition of CR streaming heating at higher CR transport moves near-disk data points off of the star forming sequence and closer to the dividing region between soft versus hard ionization.

\subsection{Other plasma diagnostics}

Figures \ref{siiv_civ}, \ref{nv_ovi}, and \ref{cii_civ} show the time evolution of the resulting line ratio profiles in height for Si-IV/C-IV, N-V/O-VI, and C-II/C-IV respectively. For Si-IV/C-IV, we plot the observed Milky Way value of $\sim 0.2$. For N-V/O-VI, we plot lines bounding the range of observed values. For C-II/C-IV, we plot the mean value from \citet[][]{fox2005multiphase}. Figure \ref{siiv_civ_scatter} \hl{ shows the mean radiative heating and $P_{\rm{cr}}/P_{\rm{th}}$, with the color representing the Si-IV/C-IV value, within the inner 10 kpc slabs of the galaxy. This plot illustrates the differences in pressure balance and radiative heating between the simulations, and what diagnostic values occur in that parameter space. Similar figures for N-V/O-IV and C-II/C-IV do not yield as much information (especially N-V/O-IV due to relatively constant profiles) on the line ratio behavior, so those figures are placed in Appendix} \ref{extra_scatter}.

Most of the simulations over predict Si-IV/C-IV, \hl{and the largest values tend to occur in $P_{\rm{cr}}$-dominated areas.} The simulation `k28' predicts  Si-IV/C-IV $\sim 10^2$, more than two orders of magnitude above the observed values in the Milky Way \citep[][]{savage2009extension, werk2019nature}. Given the lack of appropriate physical processes included in this simulation, its inability to reflect observational data is not unexpected. The addition of radiative effects in `k28\_Hrad' reduces the line ratio by a factor of 10 across the entire profile. The profile is also less variable in space and time than the profile from the `k28' run. \hl{Even though the plasma pressure is entirely dominated by CRs on average at large $H_{\rm{rad}}$, we do not see a sufficient reduction in the line ratio towards the observed values , as Figure} \ref{theoretical_line_ratio} \hl{would predict}. If we instead add CR streaming heating, as in the `k28\_Hcr' run, we see that the line ratio values are even lower, including a profile at 130 Myr that is below the observed value. The outflows in this run are more suppressed, as mentioned in Section \ref{simulation_results}, so this run does not entirely match the observed data that extends up to 15 kpc.
\hl{This simulation better isolates potential CR streaming heating behavior, as the radiative heating is not present. Interestingly, } Figure \ref{siiv_civ_scatter_hradcr} \hl{shows two regions of low line ratio values: the highest $H_{\rm{cr}}\sim 10^{-25} [\rm{erg} \rm{s}^{-1} \rm{cm^{-3}}]$ data points, and then the data points with a  slow decrease in heating at $H_{\rm{cr}}\sim 10^{-27} [\rm{erg} \rm{s}^{-1} \rm{cm^{-3}}]$.  The latter region corresponds (by comparing to the $P_{\rm{cr}}/P_{\rm{th}}$ profile) the flat line ratio profile between 2 and 8 kpc. In this region, the second most abundant phase (behind the hottest phase) is Phase 6, the intermediate gas phase.} At higher $\kappa_{\rm{cr}}$ the impact of CR streaming heating is also a suppressed wind, favoring `k29\_Hrad' over `k29\_Hrad\_Hcr'. Both runs produce Si-IV/C-IV values a factor of $\sim 5-10$ larger than the observed value. \hl{Even though `k29\_Hrad' and `k29\_Hrad\_Hcr' have different heating and pressure behavior, as } Figure \ref{siiv_civ_scatter} \hl{shows, the actual line ratio profiles are both relatively flat. Furthermore, for all simulations with CR streaming heating, significant amounts of gas exist in interval $1<P_{\rm{cr}}/P_{\rm{th}} < 3$, but we do not see the expected trend of higher CR pressure and lower line ratio values.}

The N-V/O-VI line ratio provides another metric with which to evaluate the simulations.  Values of this line ratio could favor particular underlying physical processes, such as cooling flows, shock heating, or turbulent mixing. In this work, we show as green solid lines the range of values measured observationally, as summarized by \citet[][]{wakker2012characterizing}, in order to broadly evaluate each model. The N-V and O-VI ions have a larger ionization potential than Si-IV and C-IV ions, so they trace gas phases well-above those directly influenced by radiative heating (i.e. see Figure \ref{theoretical_line_ratio}). Generally the simulations favored for N-V/O-VI are the same ones favored for Si-IV/C-IV. The simulation that least matches the observed bounds is the `k28' run, which has N-V/O-VI several orders of magnitude above the bounds. This behavior can be traced to the lack of $T \sim 10^{5-6}$ (phase 5) gas. O-VI sharply peaks within phase 5 temperature range, so a severe lack of phase 5 gas produces large line ratios. As expected by the higher ionization potentials, adding radiative heating as in `k28\_Hrad' does not significantly improve results. The runs with CR streaming heating do have predicted line values near observed bounds. At higher $\kappa_{\rm{cr}}$, both runs with and without CR streaming heating produced reasonable N-IV/O-VI values, corresponding to phase 5 gas having a significant volume filling factor in both simulations. 

C-II/C-IV traces lower temperature gas phases than the previous line ratios discussed. \hl{Similar to S-IV/C-IV, the values tend to be larger for larger $P_{\rm{cr}}$, as } Figure \ref{cii_civ_scatter} shows. As before, the `k28' over predicts the line ratio value (i.e. there is too much colder, neutral material). Adding CR streaming heating, results in profiles that are much closer to the example observations. The simulation `k28\_Hcr' (without radiative heating) in particular, produces values at 190 Myr that are quite close to the example observations, and is the only simulation at lower CR diffusivity that has an under predicting profile (120 Myr). This behavior could be due to the fact that this simulation uniquely has an outflow dominated by phase 3 gas (warm, neutral medium) as shown by the filling fraction in Figure \ref{phase_profile_grid}. Of the simulations with higher CR diffusivity, the run `k29\_Hrad\_Hcr' produces the best match to the observed line ratio values we chose.

\subsection{Caveats and further work}

In the following list we describe potential caveats in our work and additional research directions:

\begin{itemize}
    \item The observational values we use in Figures \ref{siiv_civ}, \ref{nv_ovi}, and \ref{cii_civ} are meant as rough guides to aid simulation comparison. The synthetic line ratio data points from the simulations all most certainly trace a larger range of physical situations than would be traced by a particular observation (e.g. high velocity clouds).
    \item The values of synthetic UV diagnostics depend on SFR peaks that may be fluctuations. More work is needed to see how relevant these peaks are to changes in model assumptions.
    \item We do not focus on interstellar dust effects on the predicted line ratios. 
    Dust preferentially reduces certain elements compared to others. 
    Preferential depletion of Si onto dust will lower the Si-IV/C-IV ratio allowing some previously observationally inconsistent models to be viable, as \citet[][]{werk2019nature} notes. 
    The exact depletion of elements relative to others is still debated. 
    More detailed models of dust creation/destruction \cite[e.g.][]{oberg2016photochemistry} each can influence line diagnostics based on their own resulting elemental depletion.
    \item CR transport is an active area of research. More detailed work in plasma turbulence and charged particle transport, especially at spatial scales below which the fluid approximation for CRs holds, are needed in order to better understand the impact of CRs on global galaxy scales \citep[e.g.,][]{bai2022toward}. 
    \item We use a single cooling function assuming solar metalicity. We do not explicitly track ion species with the code, and we do not include non-equilibrium effects in the cooling.
    \item Our resolution is relatively limited compared to the scale of many important processes operating at the CGM cloud pc-scales. For example, we do not resolve turbulent mixing layers, cloud formation/destruction, or CR-cloud interaction.
    \item Our model of parallel-plane radiation propagation through the domain, especially with un-resolved CGM cloud structure, is a crude approximation of real ionizing photon transport through the galaxy and halo. This work is meant to be a proof-of-concept highlighting potential UV plasma diagnostics to use as metrics to evaluate models of cosmic ray and ionizing radiation within galactic simulations. 
    \item In addition to our simulation domain being an approximation of a global galactic disk and halo, it is also not cosmological. Cosmological processes such as mergers and gas accretion are not present in our simulations.
\end{itemize}

\section{Discussion}
\label{discussion}

The BPT diagrams in Figure \ref{grid_bpt}  showed that the inner 5 kpc of our galaxy models (most comparable to the observed lines from the ISM of galaxies \citep[e.g.,][]{kewley2013cosmic}) generally follow the star forming sequence, except `k28' which explicitly excluded the stellar and metagalatic field. For some simulations, especially `k28\_Hrad', the regions farther out of the galaxy tend to cross the line between star forming and active galaxies. This is an indication that the spectra in these simulations are becoming harder with height, possibly due to higher photon absorption in the lower CGM. The fact that Figure \ref{phase_profile_grid} shows the dominant volume filling phase in `k28\_Hrad' is phase 3 (warm, neutral medium) supports this idea.

The source of the substantial reservoir of O-VI-bearing gas in the CGM, coupled with the observed N-V/O-VI and Si-IV/C-IV ratios, has been debated, especially as it related to underlying photoionzed vs. collisionally ionized gas phases. A long-lived intermediate temperature ($T \sim 10^5 - 10^6$ K) could explain these observations. In addition to several other non-equilibrium processes, CR streaming heating could also help generate this gas phase. All of our simulations with CR streaming heating produce a significant intermediate temperature volume filling phase (phases 5 and 6) in outflow regions. In fact, those phases are the dominant phase besides the hot medium (phase 7). \hl{In the simulation without radiative heating and only CR streaming heating (`k28\_Hcr'), the region approaching the observational values of Si-IV/C-IV corresponds to regions where phase 6 was significantly volume filling.} The high CR diffusivity (`k29\_Hrad') simulation did also have significant intermediate temperature gas. The exact origin of the intermediate temperature phase is unclear: is the additional thermal energy from CR collisionless heating responsible the phase, or is it that the energy loss from the CRs causes an effect (i.e. different star formation history) that eventually produces the phase? Higher CR diffusivities do lead to more stars produced, and thus more stellar feedback. 

The role of photoionization is less important than expected from Figure \ref{theoretical_line_ratio}. We hypothesize that the CR-dominated CGM pressure equilibrium would allow lower temperature phases to exist at low densities and naturally give rise to lower value for Si-IV/C-IV line ratio. \hl{In fact, the peak line ratio values in the simulations tend to occur within CR-dominated region.} The two galaxies (`k28' and `k28\_Hcr' ) with strong density overlap of the gas phases had the worst match to observed line ratio values. Photoionization did contribute during the SFR burst in `k29\_Hrad', where contribution to the Si-IV and C-IV lines tended to come from phases 3 and 4, as well as 5 in some samples. The SFR burst itself may be due to a particular feedback choice such as star formation efficiency, but regardless, the `k29\_Hrad' model is an example of how both photoionization and collisionalizion play key roles in the plasma diagnostic values. Simulations with a fixed SFR would provide additional insight into the relative importance of direct vs. indirect effects of radiative and CR feedback.

The effect of CR streaming heating on the strength of the outflow is also important. For choices of CR diffusion, streaming heating removes enough energy from the CRs such that a weaker outflow occurs, remaining below 10 kpc. At low CR diffusivity, the galaxies with collisionless heating included produced more significant quantities of intermediate temperature phases, but with the small outflow extent. The main difference between galaxies with higher CR diffusivity and those with lower CR diffusivity is the extent of the galactic outflow. One interpretation of this result is that CR confinement occurs not via self-confinement, but instead is due to extrinsic turbulence, which is not associated with collisionless plasma heating. There are many theoretical arguments regarding self-confinement versus extrinsic turbulence models that are outside the scope of this work \citep[see for example][]{yanlazarian2002,blasi2012spectral, zweibel2013microphysics, xu2021diffusion, kempski2021reconciling}.

If CR streaming heating is indeed present in the real galactic environment and assuming there are not additional wind launching mechanisms, for high $\kappa_{\rm{cr}}$, there must be a mechanism for streaming heating suppression that our model does not take into account. Our resolution scale is above that of expected pc-scale clouds of dense material in the CGM \citep[][]{gronke2018growth, farber2022survival}, so we do not model phenomena at these small scales. According to the timescale of CR losses to collisionless heating in Eq. \ref{tau_eq}, the loss timescale is proportional to $L_{\rm{cr}}$, the length scale of the cosmic ray distribution, which locally can change significantly at the cloud interface. Proper accounting of energy losses due to CR streaming heating requires accurate modeling of cloud-scale physics. The path of the CR distribution through a turbulent, clumpy medium has been an active area of interest \citep[e.g.,][]{wiener2017interaction, wiener2019cosmic, bruggen2020launching, huang2022launching}. For example, \citet[][]{bustard2021cosmic} focused on the propagation of CRs through a clumpy medium, accounting for effects of cloud partial ionization. Cloud ionization can change where the CR gradient forms at cloud interfaces, influencing where significant collisionless heating occurs. Also, partial ionization, with associated ion-neutral transport enhancement of CRs, allows CRs to more easily pass through clouds. Of course, in a more realistic model, the cloud ionization should be influenced by the stellar and metagalactic radiation field. Other wind launching mechanisms besides CRs may allow extended outflows with phase structures similar to `k29\_Hrad\_Hcr'.

The dependence of results on CR diffusivity also demonstrates the necessity of more detailed CR transport models in order to better account for the impact of feedback effects on UV diagnostics. The improvement in the values of diagnostics at low CR diffusivity could indicate that CR streaming heating indeed improves line ratios, but only accompanied by a corresponding weaker wind. In a realistic system, the CR diffusivity could vary greatly on pc-scales as the cosmic rays traverse CGM inhomogeneities: for example, \citet[][]{bustard2021cosmic} showed that the formation of CR bottlenecks depends on cloud ionization. The pressure anisotropy instability \citep[][]{zweibel2020role} may contribute to CR confinement in these environments. Other models of CR transport will similarly cause a cascade of coupled effects. For example, in the CR transport model studied in \citep[][]{holguin2019role}, CR transport is enhanced within roughly a kpc of the midplane due to turbulence, but is much less enhanced farther out of the disk where turbulence is weaker. This type of model causes enhanced star formation (and subsequent outflows powered by the stellar feedback) because CRs can more easily escape the dense disk. This model also reduces CR losses from dense gas, which leaves a greater reservoir of CR energy outside of the disk. Recent work has also investigated the potential significance of dust damping of Alfv\'en waves to \hl{CR} confinement \citep[][]{squire2021impact}. The properties of the galactic radiation field may impact the subsequent coupling of dust to the CR system.

\section{Conclusions}
\label{conclusions}

\hl{The scope of this work is to provide a proof-of-concept study of the interaction between two complex galactic feedback mechanisms, ionizing radiation and CRs, and the resulting effect on UV diagnostics. We sketch out how such an analysis may be conducted and highlight where complex feedback interactions may or may not occur. More detailed modeling of feedback mechanisms, galactic outflows/inflows, and methods for calculating synthetic diagnostics will reveal a wealth of information about the galactic and circumgalactic environment.}

We perform MHD simulations of a section of a galactic disk, including radiative heating and CR feedback. We test the effects of omitting radiative heating, the value of CR transport parallel to the magnetic field ($\kappa_{\rm{cr}} = 10^{28}$ or $10^{29} \ \rm{cm}^2 s^{-1}$), and the inclusion of collisionless CR streaming heating. We post-process the simulations using parallel-plane Cloudy models informed by the simulation results in order to produce synthetic observations of four UV diagnostics: the BPT diagram, Si-IV/C-IV, N-V/O-VI, and C-II/C-IV. We compare the gas phase structure, predicted UV diagnostics compared to observational values,  and evaluate the relative ability of the simulations to reproduce observations. 

Our conclusions are:

\begin{itemize}
    \item \hl{We demonstrate that including radiative heating in simulations can have a substantial impact on observable Si-IV/C-IV values, while unable to improve N-V/O-VI values (due to the higher ionization potential). Although, this effect suffers some interdependence among other model parameters that is challenging to quantify.  Nonetheless, variations of other model parameters, like star formation history, are likely playing a sub-dominant role. Other feedback mechanisms may be needed to better simultaneously match a variety of UV diagnostics.}
    \item The inclusion of CR streaming heating in our simulations reduces the strength of galactic outflows due to the additional energy losses of the CR population.
    \item In simulations with slower CR transport, the addition of CR streaming heating results in synthetic diagnostics values closer to observed values due to the production of intermediate, transition ($T \sim 10^{5-6}$) temperature gas. This finding lends credibility that this intermediate phase can exist within a CR-driven outflow. However, slower CR transport results in a less extended outflow.
    \item The simulations with the inclusion of faster CR transport produce the closest predictions to observed UV diagnostics due to the production of transitional temperature gas. Without CR streaming heating losses included in feedback, the outflows extend out to the edge of the domain at a height of 20 kpc above the midplane. 
    \item The type of analysis presented in this proof-of-concept also demonstrates the potential use of UV diagnostics as additional constraints on key galactic physical models that are currently not well understood, such as cosmic ray transport and other feedback mechanisms. 
\end{itemize}

\section{Acknowledgements}

We thank Yakov Faerman, Max Gronke, Dongwook Lee, Zhijie Qu, Nicolas Trueba, and Mateusz Ruszkowski for useful conversations. F.H. acknowledges support from the NASA FINESST fellowship (grant number 80NSSC20K1541), and the University of Michigan Rackham Predoctoral fellowship. R.J.F. thanks the Max Planck Society for support through the Max Planck Research Group of Dr. Max Grönke, the Multiphase Gas Group. J.K.W. acknowledges support from an NSF-CAREER, award number 2044303. Computational resources for this work were provided by the Texas Advanced Supercomputer Center (TACC) Stampede2 machine (via XSEDE), NASA High-End Computing Programming on the Pleiades machine at NASA Ames Research Center, and some of the simulations and analysis in this work have been supported by the Max Planck Computing and Data Facility (MPCDF) computer cluster Freya. Data analysis presented in this paper was performed with the publicly available \emph{yt} visualization software \citep{turk2011}. We are grateful to the \emph{yt} community and development team for their support.

\section{Data Availability}
The data and code underlying this article will be shared on a reasonable request to the corresponding author.

\bibliographystyle{mnras}
\bibliography{main}

\appendix

\section{Radiative heating and pressure, with associated plasma diagnostic values}
\label{extra_scatter}
\begin{figure*}
    \centering
    \includegraphics[width=0.9\textwidth]{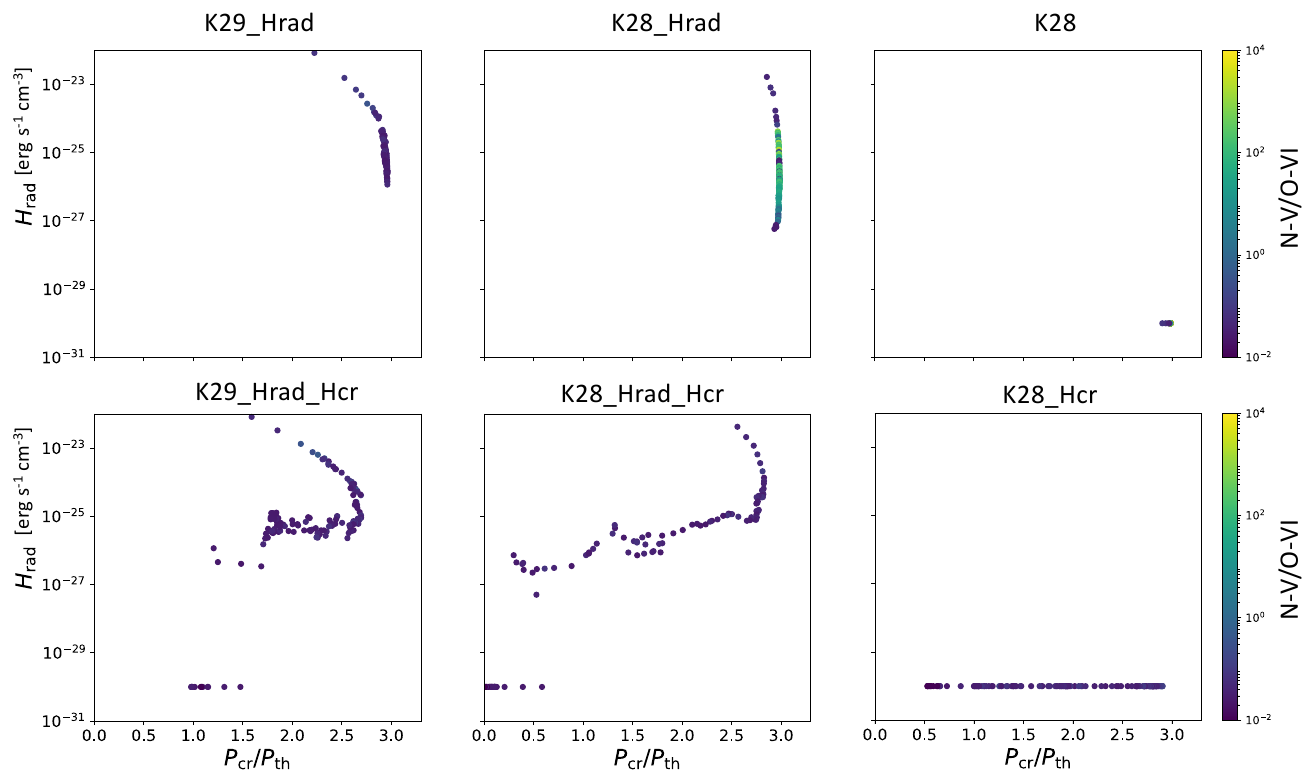}
    \caption{\hl{Similar figure to Figure }\ref{siiv_civ_scatter}, \hl{except for the N-V/O-VI line ratio.}}
    \label{nv_ovi_scatter}
\end{figure*}

\begin{figure*}
    \centering
    \includegraphics[width=0.9\textwidth]{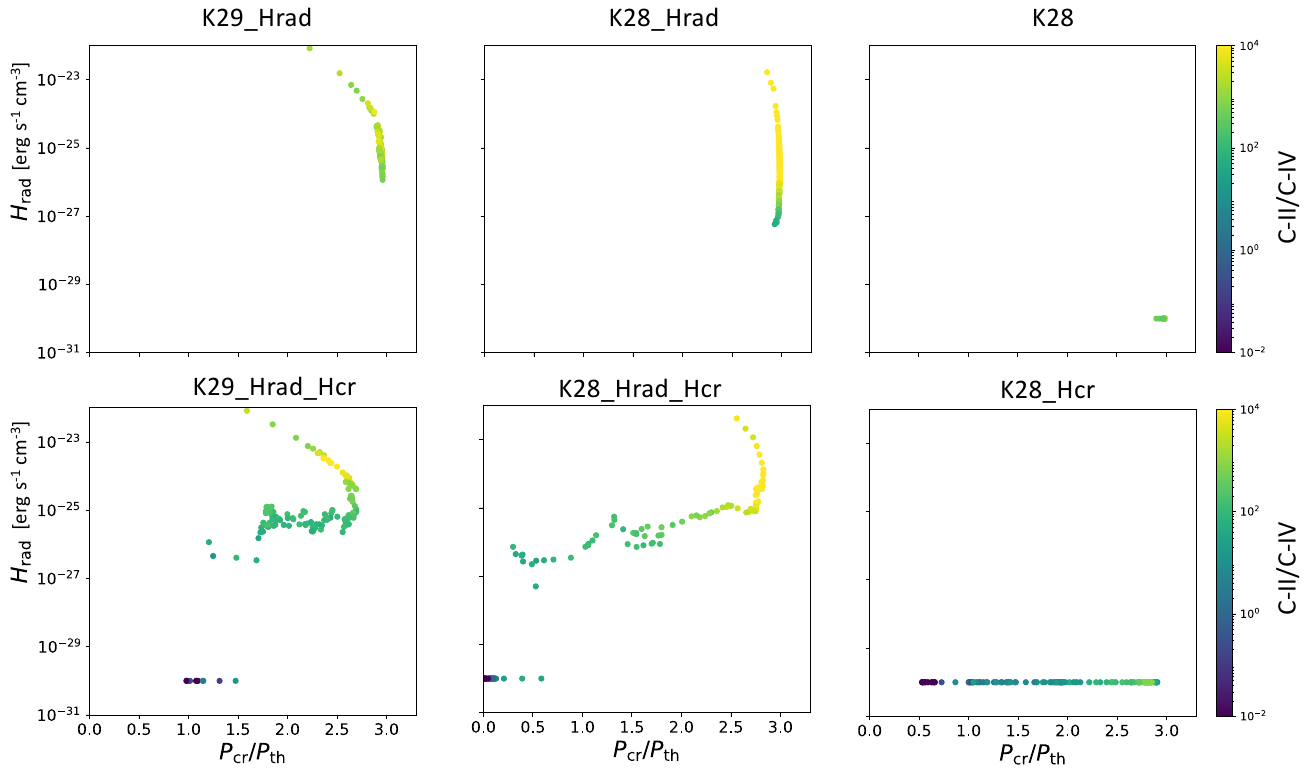}
    \caption{\hl{Similar figure to Figure }\ref{siiv_civ_scatter}, \hl{except for the C-II/C-IV line ratio.} }
    \label{cii_civ_scatter}
\end{figure*}

\begin{figure*}
    \centering
    \includegraphics[width=0.9\textwidth]{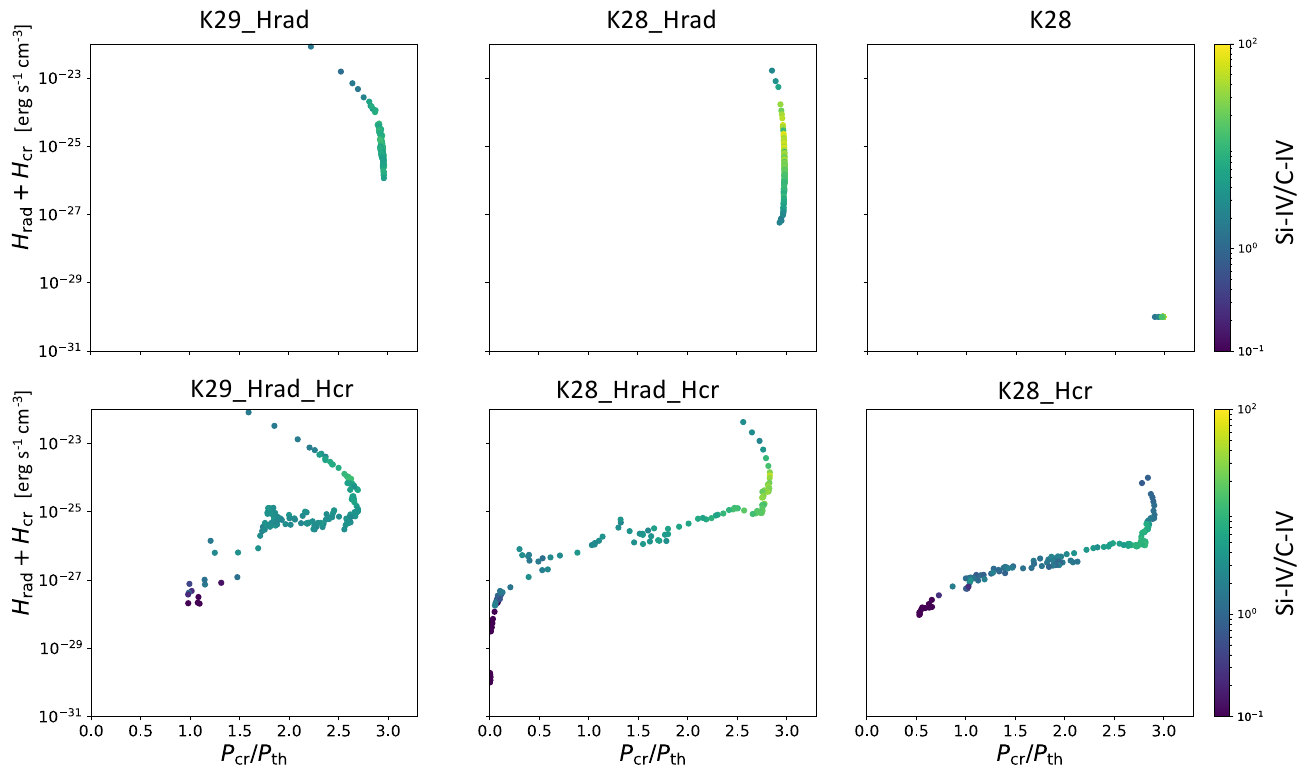}
    \caption{\hl{Similar figure to Figure }\ref{siiv_civ_scatter}, \hl{except the y-axis is $H_{\rm{rad}} + H_{\rm{cr}}$. } }
    \label{siiv_civ_scatter_hradcr}
\end{figure*}

This appendix contains additional plots that show the gas properties in galaxy and the diagnostic line ratios. Figures \ref{nv_ovi_scatter} and \ref{cii_civ_scatter} \hl{plot the N-V/O-VI and C-II/C-IV values on top of the same axes as Figure} \ref{siiv_civ_scatter}. Figure \ref{siiv_civ_scatter_hradcr} \hl{adds collisionless CR streaming to the total radiative heating rate. The general result on the plots is that the regions with minimal radiative heating seen in Figure} \ref{siiv_civ_scatter} \hl{no longer have the floor value of the heating rate. These data points are at larger midplane heights. The difference is most noticeable in `k28\_Hcr', since there is not radiative heating included.}

\end{document}